\documentclass[10pt]{article}
\usepackage[dvips]{graphicx}
\usepackage{emlines2}
\newcommand{\bmp}[1]{\begin{minipage}[c]{#1}}
\newcommand{\emp}{\end{minipage}}
\newcommand{\be}{\begin{equation}}
\newcommand{\ee}{\end{equation}}
\newcommand{\een}{\end{enumerate}}
\newcommand{\ben}{\begin{enumerate}}
\newcommand{\ber}{\begin{eqnarray}}
\newcommand{\eer}{\end{eqnarray}}

\newcommand{\nn}{\nonumber}

\newcommand{\bp}{\mbox{\boldmath$p$}}
\newcommand{\bk}{\mbox{\boldmath$k$}}

\newcommand{\M}{{\cal M}}
\newcommand{\ep}{\varepsilon}

\begin{document}
\hfill PRA-HEP 97/15

\hfill October 2, 1997\\[8ex]
\begin{center}
{\Large\bf Vector Boson Scattering in the Standard Model -- \\[0.2cm]
                 an Overview of Formulae\\[3ex]}
Tom\'{a}\v{s} Bahn\'{\i}k\footnote{e-mail: tomas.bahnik@vslib.cz}\\[2ex]
{\em Department of Physics, Technical University Liberec,\\
H\'{a}lkova 6, 461 17 Liberec, Czech republic\\[2ex]}
and\\[2ex]
{\em Nuclear Centre, Faculty of Mathematics and Physics,\\
Charles University, V Hole\v{s}ovi\v{c}k\'{a}ch
2, 180 00 Prague 8, Czech republic\\[5ex]}
\end{center}
\begin{abstract}
Tree-level scattering amplitudes of longitudinally polarized electroweak
vector bosons in the Standard Model are calculated using {\em Mathematica}
package {\em Feyncalc}. The modifications of low-energy theorems for
longitudinally polarized $W$ and $Z$ in the Standard Model are
discussed.
\end{abstract}
\section{Introduction}
One of the open questions in high energy physics is the mechanism of
electroweak symmetry breaking (EWSB). The physics that breaks
electroweak symmetry is responsible for giving the $W$ and $Z$ their
masses. Since a massive spin-one particle has three polarizations,
rather than the two of a massless mode, the new physics must supply
degrees of freedom to be swallowed by the $W$ and $Z$. These new degrees
of freedom are the longitudinal polarizations $W_L$ and $Z_L$ of the
vector bosons. Therefore the interactions of the longitudinal components
of the vector bosons could provide a good way to probe the interactions
of the symmetry breaking sector \cite{gaillard}.

The interactions of $W_L$ and $Z_L$ in high energy processes are usually
studied by means of different production mechanisms of vector bosons
followed by their purely leptonic decays  {\em e.g.} $W^{\pm} \to
l^{\pm}\nu$ and $Z\to l^+l^- (l=e,\mu)$, referred to as gold-plated
channels. One of the production  mechanism is through light fermion
anti-fermion {\em i.e.} $q\bar{q}$ or $e^+ e^-$ annihilation. This
yields vector boson pairs that are mostly transversely polarized and is
usually a background to the other processes. The important exception is
the production of longitudinally polarized vector bosons through new
vector resonance.

A second mechanism for producing longitudinal vector boson pairs in
hadron colliders is gluon fusion. The initial gluons turn into two
vector bosons via an intermediate state that couples to both gluons and
electroweak vector bosons like the top quark or new colored particles of
a technicolor model \cite{jbagger}.

Finally, there is the vector-boson fusion process when vector bosons are
radiated by colliding fermions and then rescattered. When the fermions
are quarks then the process of vector boson scattering is considered as
a subprocess of subprocess in hadron collision. Sensitivity of different
types of colliders to the above mentioned processes has been discussed
in a series of articles \cite{SISBS}.

In this paper I calculate exact tree-level scattering amplitudes of
longitudinally polarized electroweak vector bosons in the Standard Model.
The aim of the present paper is to check independently the existing
results \cite{dutta,bento,barger,duncan}, in particular because I think
there has been a minor error in an earlier paper \cite{bento}. As a
consistency test I use the high-energy ($E\gg m_W$) behaviour of a
pure gauge amplitudes. Their quadratic growth ($\sim E^2$)
should be canceled by introducing Standard Model Higgs boson.

It has been shown that some universal low-energy theorems (LET) for
the scattering of longitudinally polarized $W$ and $Z$ hold \cite{LET}.
These theorems are valid below the scale $\Lambda_{SB} \sim\,$1\,TeV,
provided that the symmetry breaking sector contains no particles much
lighter than $\Lambda_{SB}$. The derivation in \cite{LET} shows that in
this case the LET are given by $SU(2)_L \times U(1)_Y$ gauge interactions
of vector bosons alone. The particle which modifies pure gauge
amplitudes (and LET) in the SM is the Higgs boson. To see this
modifications, I plot the complete amplitudes for different values of Higgs
boson mass and compare them with the pure gauge contributions.

\section{Scattering Amplitudes}
Calculation of the scattering amplitudes of the gauge bosons by hand is
a tedious task. The use of function {\tt SpecificPolarization} of the
{\em Mathematica} package {\em FeynCalc} \cite{mertig} has substantially
reduced algebraical manipulations.

Longitudinal polarization picks up a specific direction in space.
Thus the amplitudes written in terms of Mandelstam variables $s,t,u$
do not describe scattering of {\em longitudinally}  polarized
particles in {\em all} Lorentz frames. For the process
$V(p_1) + V(p_2)\to V(k_1) + V(k_2)$ the function
{\tt SpecificPolarization} uses the following representation of the
longitudinal polarization vectors $\varepsilon (p,L)$
\ber
   \varepsilon^{\mu}(p_1,L) &=& F^{\mu }_{L}(p_1,\,k_1,k_2)\nn\\
   \varepsilon^{\mu}(p_2,L) &=& F^{\mu }_{L}(p_2,\,k_1,k_2)\nn\\
   \varepsilon^{\mu}(k_1,L) &=& F^{\mu }_{L}(k_1,\,p_1,p_2)\nn\\
   \varepsilon^{\mu}(k_2,L) &=& F^{\mu }_{L}(k_2,\,p_1,p_2)
\label{longpolvec}
\eer
where
\be
\nn F_{L}^{\mu }(r,\,a,b) = \frac{ r^{\mu}(b\cdot r+a\cdot r) -
   (a+b)^{\mu }r^{2}}{\sqrt{r^{2}[(b\cdot r+a\cdot r)^{2}-
   r^{2}(b+a)^{2}]}}
\ee
Using this function, we get the longitudinal polarization only in
the CM system, where $\bp_1 + \bp_2 = \bk_1 + \bk_2 = 0$.

Table\,\ref{tab:indproc} summarizes tree-level contributions (contact
graphs are not listed) to the scattering amplitudes of the gauge bosons
in the SM in the $U$-gauge.
\begin{table}[h]
\begin{center}
\begin{tabular}{|c|c|c|c|c|c|}\hline
process \# & process & $s$ & $t$ & $u$ & references \\ \hline\hline
\rule[-3mm]{0cm}{8mm} 1 & $W^+_1 W^-_2 \to Z_3 Z_4$ & $H$ & $W$ & $W$ &
\cite{dutta,bento}\\ \hline
\rule[-3mm]{0cm}{8mm} 2 & $W^+_1 Z_2 \to Z_3 W^+_4$ & $W$ & $W$ & $H$ &
\cite{bento}
  \\ \hline
\rule[-3mm]{0cm}{8mm} 3 & $W^+_1 W^+_2 \to W^+_3 W^+_4$ &  &
$Z, \gamma, H$ & $Z, \gamma, H$ & \cite{barger} \\ \hline
\rule[-3mm]{0cm}{8mm} 4 & $W^+_1 W^-_2 \to W^+_3 W^-_4$ & $Z, \gamma, H$
& $Z, \gamma, H$ & & \cite{duncan} \\ \hline
\end{tabular}
\caption{\small Individual processes together with exchanged particles in
different channels ($s, t, u$). $W, Z, \gamma$ are electroweak gauge bosons and
photon and $H$ is the SM Higgs boson. Contact graphs contribute to
each of the process and are not listed.
\label{tab:indproc}}
\end{center}
\end{table}
Contributions to the process \#$n$ from graph with  $V$-boson exchange in the
$k$-channel  are denoted as
\be \M^{(n)}_{kV} = -\frac{g_1 g_2}{k - m_V^2} A^{(n)}_{k} \label{mnkv} \ee
and from the contact graph
\be \M^{(n)}_c = g\, A^{(n)}_c  \label{mnc} \ee
with $g_1, g_2$ and $g$ denoting relevant coupling constants. Amplitudes
given by the low-energy theorems are denoted as $\M^{(n)}_{LET}$. Note
that in the paper of Bento and Llewellyn Smith \cite{bento} the $WZ \to
WZ$ amplitude is claimed to be (in the current notation) \[\M^{(2)} =
\M^{(2)}_c+\M^{(2)}_{uH}+\M^{(2)}_{uW}+\M^{(2)}_{tW}\qquad\mbox{(wrong)}\]
and is also calculated in this way. This is wrong, because $W$ is
exchanged in $s$ and $t$ (or $u$) channel\footnote{There is the
interchange $t \leftrightarrow u$ in \cite{bento} in comparison with
this paper.}.

Explicit formulae for the scattering amplitudes are somewhat difficult to
read and are postponed to the appendix\,\ref{exactformulae}.
Table\,\ref{tab:relA} summarizes
relations among different $A$ parts of gauge amplitudes as defined in
(\ref{mnkv}) and (\ref{mnc}). These relation are almost obvious, nevertheless
they can be used as a check.
\begin{table}[h]
\begin{center}
\begin{tabular}{|c|c|c|}\hline
process \# & Gauge boson exchange & Contact graph  \\ \hline\hline
\rule[-3mm]{0cm}{8mm} 1 & $A^{(1)}_{tW} (\cos{\theta_{cm}}) =
A^{(1)}_{uW} (- \cos{\theta_{cm}})$
 & $A^{(1)}_c $  \\ \hline
\rule[-3mm]{0cm}{8mm} 2 & $A^{(2)}_{sW},\ A^{(2)}_{tW}$
  &$A^{(2)}_c$  \\ \hline
\rule[-3mm]{0cm}{8mm} 3 & $A^{(3)}_{(t,u)(\gamma,Z)} = -
A^{(1)}_{(t,u)W}\big|_{m_Z=m_W}$ & $A^{(3)}_c =
A^{(1)}_c\big|_{m_Z=m_W}$  \\ \hline
\rule[-3mm]{0cm}{8mm} 4 & $A^{(4)}_{(s,t)(\gamma,Z)} = -
A^{(2)}_{(s,t)W}\big|_{m_Z=m_W}$ & $A^{(4)}_c = A^{(2)}_c\big|_{m_Z=m_W}$
 \\ \hline
\end{tabular}
\caption{\small Relations among the $A$ parts of the scattering
amplitudes.
\label{tab:relA}}
\end{center}
\end{table}
Besides these relations we have  $A^{(1)}_{tW}\big|_{m_Z=m_W} = -
A^{(2)}_{tW}\big|_{m_Z=m_W}$ but $A^{(2)}_{tW} \neq - A^{(1)}_{tW}$.
Note that the $A^{(n)}_{\gamma} = A^{(n)}_{Z}$ because the longitudinal
term in propagator ($\sim q^{\alpha} q^{\beta}$) does not contribute to
the amplitudes with exchange of neutral gauge boson. 

\subsection{$W^+(k_1) +  W^-(k_2) \to Z(k_3)+ Z(k_4)$}
On the tree-level the pure gauge contributions are from $W$-exchange in $t$ and $u$
channels and the contact graph
\be  \M^{(1)}_{kW}= \frac{-  g^2\cos^2{\theta_W}}{k - m_W^2}
  A^{(1)}_{kW}\qquad k=t,u  \ee
\be  \M^{(1)}_c= -  g^2\cos^2{\theta_W} A^{(1)}_c \ee
Complete gauge amplitude grows linearly with $s$
\ber \M^{(1)}_{gauge} &=&
\M^{(1)}_{tW} + \M^{(1)}_{uW} + \M^{(1)}_c \nn\\
&=& \M^{(1)}_{LET} - g^2\frac{(1-x^2) - 2 \cos^2{\theta_W} \rho (1+x^2)}
{2 \rho^2 \cos^2{\theta_W}(1-x^2)} +  O\left(\frac{m_W^2}{s}\right)
\label{m1gauge}
\eer
where
\[\M^{(1)}_{LET} = \frac{g^2\,s}{4\,\rho m_W^2}, \quad \rho =
\frac{m_W^2}{m_Z^2 \cos^2{\theta_W}} \quad\mbox{and}\quad\ x \equiv
\cos{\theta_{cm}}.\]
with $\theta_{cm}$ an angle between $\bk_1$ and $\bk_3$ in  CMS.

In the SM this growth is canceled by exchange of the Higgs boson in
the $s$-channel
\be \M^{(1)}_{sH} = - \frac{g^2 m_W m_Z}{\cos{\theta_W}}
    \frac{(\ep_1\cdot\ep_2) (\ep^*_3\cdot\ep^*_4)}{s - m_H^2 + i m_H\Gamma_H}
\label{m1sh}
\ee
The amplitude $\M^{(1)}_{sH}$ has, for longitudinal polarizations, the
high-energy  ($E\gg m_H, m_W$) expansion (for exact formulae see appendix)
\[\M^{(1)}_{sH} = - \frac{g^2 m_W m_Z}{\cos{\theta_W}} \left [ \frac{s}{4 m_Z^2 m_W^2}
+ O(s^0) \right ] = -\frac{g^2\sqrt{\rho}}{4 m_W^2}\,s + O(s^0)\]
so the cancellation occurs for $\rho = 1$.

\begin{figure}[t]
\begin{center}
\includegraphics[height=8cm,width=10cm]{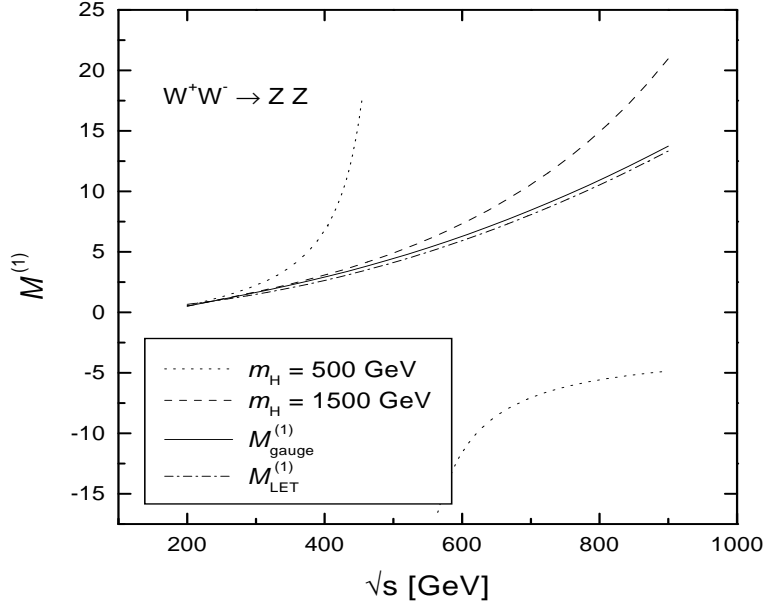}
\end{center}
\caption{\small Tree-level amplitudes of the process $WW\to ZZ$
as a function of $\sqrt{s}$ for different values of $m_H$. The Higgs
boson width, $\Gamma_H$, is set to zero, $\cos{\theta_{cm}} = 0.5$,
$\rho=1$, $m_Z=91.2$\,GeV, $\sin^2{\theta_W} = 0.231$,
$g^2=\frac{4\pi\alpha_{em}}{\sin^2{\theta_W}}\doteq 0.42$.
\label{figwwzz}}
\end{figure}

Figure\,\ref{figwwzz} shows dependence of the complete amplitude $\M^{(1)} =
\M^{(1)}_{gauge} + \M^{(1)}_{sH}$ on the Higgs boson mass, $m_H$, in the
region $m_W < \sqrt{s} < 1\,{\rm TeV}$ and compares it with
$\M^{(1)}_{gauge}$ and $\M^{(1)}_{LET}$.
Note that $\M^{(1)}_{gauge}$ differs from $\M^{(1)}_{LET}$ in the limit $s
\to \infty$ by a constant term. Numerical values of all parameters are
the same throughout the paper.

\subsection{$W^+(k_1)+ Z(k_2) \to Z(k_3) + W^+(k_4)$}
\begin{figure}[t]
\begin{center}
\includegraphics[height=8cm,width=10cm]{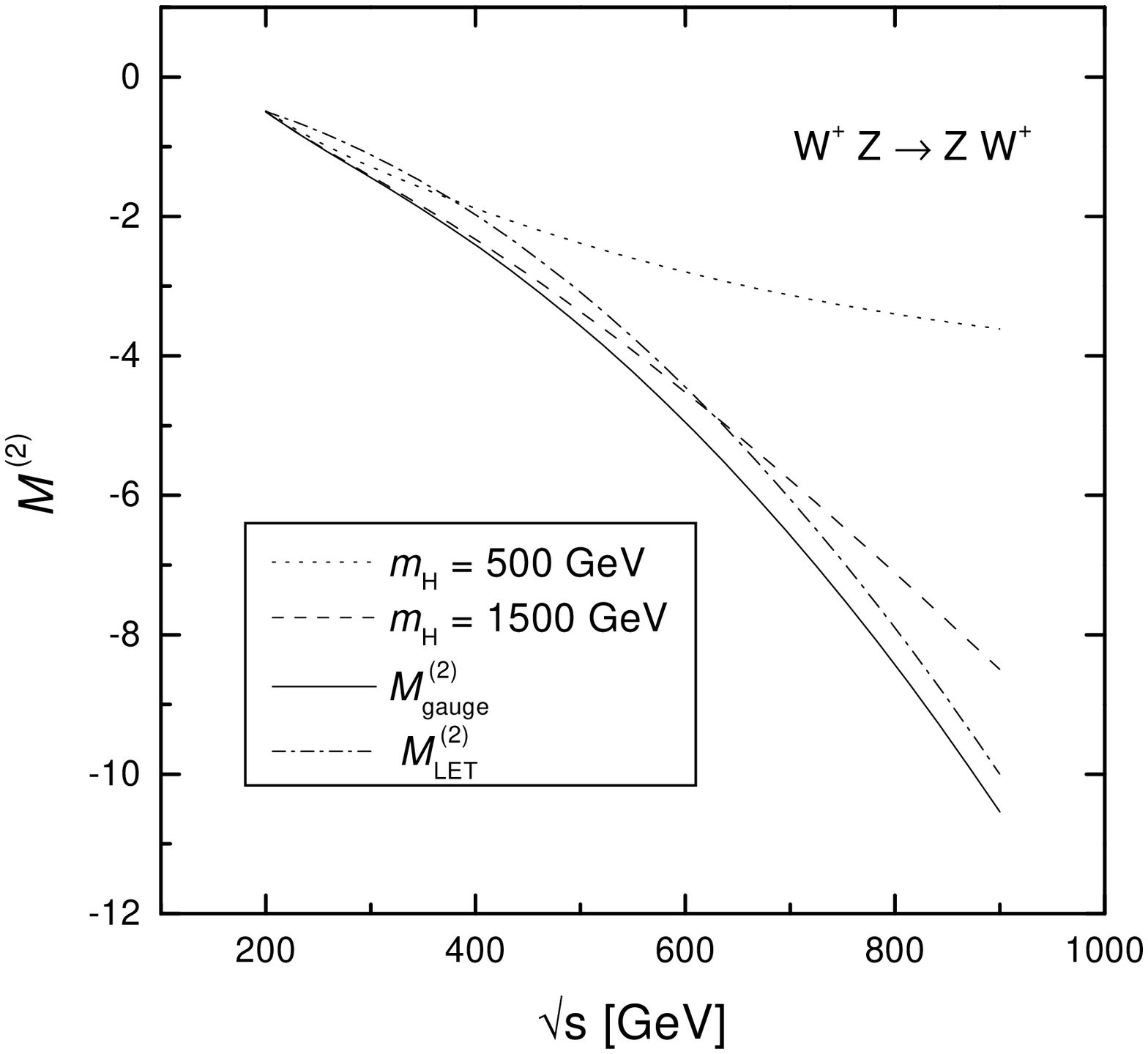}
\end{center}
\caption{\small Tree-level amplitudes of
the process $WZ\to WZ$.
\label{figwzwz}}
\end{figure}
On the tree-level the pure gauge contributions are from $W$-exchange in
$s$ and $t$
channels and the contact graph
\be  \M^{(2)}_{kW}= \frac{-  g^2\cos^2{\theta_W}}{k - m_W^2}
A^{(2)}_{kW} \quad k = s, t \ee
\be  \M^{(2)}_c= -  g^2\cos^2{\theta_W} A^{(2)}_c \ee
The asymptotic behaviour of the complete tree-level gauge amplitude is
\ber
\M^{(2)}_{gauge} &=& \M^{(2)}_{sW} + \M^{(2)}_{tW} + \M^{(2)}_c \nn\\
&=& \M^{(2)}_{LET} -\frac{g^2 (2x(1-x) +
\rho\cos^2{\theta_W}(3+2x-x^2))}{4 \rho^2\cos^2{\theta_W}
(1-x)} + O\left(\frac{m_W^2}{s}\right)
\label{m2gauge}
\eer
where a low-energy amplitude is usually defined as
\[\M^{(2)}_{LET} =  \frac{g^2 u}{4\rho m_W^2}=
-\frac{g^2 s}{8\rho m_W^2}\,(1+\cos{\theta_{cm}})+O(s^0)\ .
\]
The exchange of the Higgs boson in the $u$-channel
\be
\M^{(2)}_{uH} = - g^2 \frac{m_W m_Z}{\cos{\theta_W}}
\frac{(\ep_1\cdot\ep^*_4)\,(\ep_2\cdot\ep^*_3)}{u - M^2_H}
\label{m2uh}
\ee
has for longitudinal polarizations the high-energy expansion
\[ \M^{(2)}_{uH} =
\frac{g^2 \sqrt{\rho}}{8 m_W^2} \,(1+\cos{\theta_{cm}})\,s +O(s^0)=
- \frac{g^2 \sqrt{\rho}}{4 m_W^2}\,u + O(s^0)\ .
\]
Figure\,\ref{figwzwz} shows complete tree-level amplitude $\M^{(2)}$
for different $m_H$ and compares it with $\M^{(2)}_{gauge}$ and
$\M^{(2)}_{LET}$.

\subsection{$W^+(k_1) + W^+(k_2) \to W^+(k_3) + W^+(k_4)$}
\begin{figure}[t]
\begin{center}
\includegraphics[height=8cm,width=10cm]{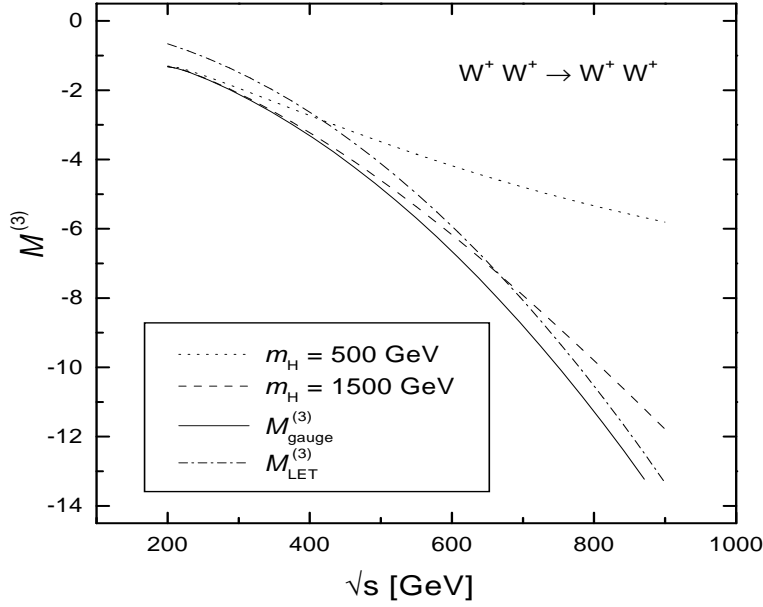}
\end{center}
\caption{\small Tree-level amplitudes of the process $W^+W^+\to
W^+W^+$.
\label{figwpwp}}
\end{figure}
On the tree-level graphs with $\gamma$ and $Z$ exchange in $t$ and $u$
channels and contact graph contribute. In the Standard Model,
Higgs boson exchange in $t$ and $u$
channels gives the desired high energy behaviour.
\[\M^{(3)}_{kZ}= \frac{-g^2\cos^2{\theta_W}}{k - m_Z^2}
A^{(3)}_{kZ}\qquad
\M^{(3)}_{k\gamma}= \frac{-e^2}{k} A^{(3)}_{k\gamma} \qquad k = t,u\ .\]
Because longitudinal term in the $Z$ boson propagator does not
contribute we have
\[A^{(3)}_{k\gamma} = A^{(3)}_{kZ} \ .\]
Contact graph amplitude for longitudinally polarized gauge bosons has
the form
\be \M^{(3)}_c = g^2 A^{(3)}_c = \frac{g^2 s}{8\,m_W^4}\,
(-8\,m_W^2 + 3\,s - s\,\cos^2{\theta_{cm}})
\ee
The gauge amplitude can be written as
\be \M^{(3)}_{gauge} = - g^2\cos^2{\theta_W}\left[\frac{A^{(3)}_{tZ}}{t -
m_Z^2} + \frac{A^{(3)}_{uZ}}{u - m_Z^2}\right] - g^2\sin^2{\theta_W}
\left[\frac{A^{(3)}_{t\gamma}}{t} + \frac{A^{(3)}_{u\gamma}}{u}\right] +
\M^{(3)}_c
\label{m3gauge}
\ee
Expanding this expression in powers of $s$ gives
\ber \M^{(3)}_{gauge} = &-& g^2\cos^2{\theta_W}\left[\frac{3-\cos^2{\theta_{cm}}}{8 m_W^4}\,s^2
- \frac{3 m_Z^2}{4 m_W^4}\,s + O(s^0)\right]\nn \\
&-&
g^2\sin^2{\theta_W}\,\left[\frac{3-\cos^2{\theta_{cm}}}{8 m_W^4}\,s^2 +O(s^0)\right] \nn \\
&+& g^2 \left[\frac{3-\cos^2{\theta_{cm}}}{8 m_W^4}\,s^2 - \frac{s}{m_W^2}\right]
\eer
Quadratic (in $s$) divergencies are canceled and including constant
terms of the order $O(s^0)$ we get
\be
\M^{(3)}_{gauge} = \M^{(3)}_{LET}- g^2\,\frac{
\left( 3 + {x^2} - \rho\,\cos^2{\theta_W} (6 -
       4\,{\rho} + 10{x^2} - 12\,{\rho}\,{x^2}\right)}
   {2\,{\rho^2}\,\left( 1 - {x^2} \right) \,{\cos^2{\theta_W}^2}} +
   O\left(\frac{m_W^2}{s}\right)
\label{asym3G}
\ee
where
\be
\M^{(3)}_{LET} = - \frac{g^2\,s}{4 m_W^2}\left(4 -
\frac{3}{\rho}\right) \ .
\label{let3}
\ee
This linear divergence should be canceled by Higgs exchange in $t$ and
$u$ channels
\be
\M^{(3)}_{H} = - g^2 m_W^2 \left[\frac{(\ep_1\cdot\ep^*_3)
(\ep_2\cdot\ep^*_4)}{t-m_H^2} +\frac{(\ep_1\cdot\ep^*_4)
(\ep_2\cdot\ep^*_3)}{u - m_H^2}\right]
\label{m3h}
\ee
with high-energy expansion
\be \M^{(3)}_{H} = \frac{g^2\,s}{4 m_W^2} + O(s^0) \label{asym3H} \ .\ee
Comparing (\ref{let3}) and (\ref{asym3H}) we see that $\rho = 1$
ensures desired cancellation.

\subsection{$W^+(k_1)+ W^-(k_2) \to W^+(k_3)+ W^-(k_4)$}
\begin{figure}[t]
\begin{center}
\includegraphics[height=8cm,width=10cm]{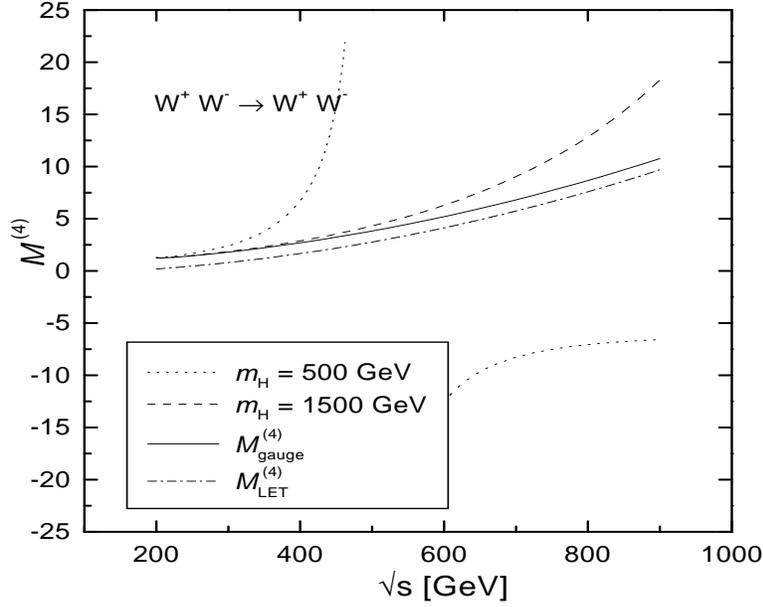}
\end{center}
\caption{\small Tree-level amplitudes of the process $W^+W^-\to
W^+W^-$.
\label{figwpwm}}
\end{figure}
In this case we have to consider  $Z$, $\gamma$ and Higgs boson exchange
in  $s$ and $t$ channels and contact graph.
As in the previous sections we write
\[\M^{(4)}_{kZ}= \frac{-g^2\cos^2{\theta_W}}{k - m_Z^2}
A^{(4)}_{kZ} \qquad \M^{(4)}_{k\gamma}= \frac{-e^2}{k}
A^{(4)}_{k\gamma}\qquad k= s,t\]
The results for the $A$
parts of the amplitude can be obtained directly from corresponding
formulae for the process $W^+ Z \to W ^+ Z$ by setting $m_Z = m_W$
\[A^{(4)}_{(s,t)(Z,\gamma)} = - A^{(2)}_{(s,t)W}\big|_{m_Z = m_W}\]
or we can also notice that $A^{(4)}_{tZ} = - A^{(3)}_{tZ}$. Again we
have $A^{(4)}_{kZ} =  A^{(4)}_{k\gamma}$.
In the case of longitudinally polarized gauge bosons
\be \M^{(4)}_c = g^2 A^{(4)}_c = \frac{g^2 s}{16\,m_W^4}\,(8\,m_W^2 - 3\,s - 24\,m_W^2\,x
+ 6\,s\,x + s\,x^2)
\ee
Let us examine high-energy expansion of gauge amplitude
\be \M^{(4)}_{gauge} = -g^2\cos^2{\theta_W}\left[\frac{A^{(4)}_{sZ}}{s-m_Z^2}
+ \frac{A^{(4)}_{tZ}}{t-m_Z^2}\right] -
g^2\sin^2{\theta_W}\left[\frac{A^{(4)}_{s\gamma}}{s} +
\frac{A^{(4)}_{t\gamma}}{t}\right] + \M^{(4)}_c \ .
\label{m4gauge}
\ee

\ber \M^{(4)}_{gauge} = - &g^2 \cos^2{\theta_W}&\left(\frac{x}{4
m_W^4}s^2 + \frac{m_Z^2 x}{4 m_W^4} s + \frac{x^2 +2x -3}{16 m_W^4} s^2 +
\frac{3 m_Z^2 -16 m_W^2 x + m_Z^2 x}{8 m_W^4} s\right)\nn\\
- &g^2 \sin^2{\theta_W}&\left(\frac{x}{4m_W^4}s^2 + \frac{x^2 +2x -3}{16
m_W^4} s^2 - \frac{2 x}{m_W^2} s\right) + O(s^0) \nn\\
+ &g^2& \left(\frac{x^2 + 6x -3}{16 m_W^4} s^2 +\frac{1 - 3x}{2 m_W^2}s
\right)
\eer
After simplification and including terms of order $O(s^0)$ we get
\ber \M^{(4)}_{gauge} &=& \M^{(4)}_{LET}\nn\\ &+&
g^2 \frac{(3 + x^2 - \rho\,\cos^2{\theta_W}(12 -
        12\,\rho + 12\,x - 16\,\rho\,x - 8\,x^2 +
        12\,\rho\,x^2))}
        {4\,\rho^2\,( 1 - x)\,\cos^2{\theta_W}}\nn\\&+&
        O\left(\frac{m_W^2}{s}\right)\nn
\eer
where as in the case of the process \#1 I denote
\[\M^{(4)}_{LET} = -\frac{g^2 u}{4m_W^2}\left(4-\frac{3}{\rho}\right)\]
in accordance with \cite{LET}.
High-energy expansion of the Higgs boson contribution
\be
\M^{(4)}_{H} = -g^2 M^2_W\left[\frac{(\ep_1\cdot\ep_2) (\ep_3^*\cdot\ep_4^*)}
{s-m_H^2} + \frac{(\ep_1\cdot\ep_3^*) (\ep_2^*\cdot\ep_4^*)}{t-m_H^2}\right]
\label{m4h}
\ee
is
\[\M^{(4)}_{H} =
 - \frac{g^2 s}{8 m_W^2} (1 + \cos{\theta_{cm}}) + O(s^0) =
\frac{g^2 u}{4 m_W^2} + O(s^0)
\]
\begin{appendix}
\section{Exact formulae \label{exactformulae}}
The following Feynman rules and notation is used (all momenta are
outgoing) \cite{horejsi}
\\[1cm]
\bmp{3.5cm}
\special{em:linewidth 0.3pt}
\unitlength 0.15mm
\linethickness{0.3pt}
\begin{picture}(171.00,174.00)
\emline{10.00}{10.00}{1}{12.35}{9.93}{2}
\emline{12.35}{9.93}{3}{14.38}{10.17}{4}
\emline{14.38}{10.17}{5}{16.11}{10.71}{6}
\emline{16.11}{10.71}{7}{17.52}{11.57}{8}
\emline{17.52}{11.57}{9}{18.63}{12.74}{10}
\emline{18.63}{12.74}{11}{19.42}{14.21}{12}
\emline{19.42}{14.21}{13}{19.91}{16.00}{14}
\emline{19.91}{16.00}{15}{20.00}{20.00}{16}
\emline{20.00}{20.00}{17}{19.94}{22.13}{18}
\emline{19.94}{22.13}{19}{20.25}{24.00}{20}
\emline{20.25}{24.00}{21}{20.94}{25.63}{22}
\emline{20.94}{25.63}{23}{22.00}{27.00}{24}
\emline{22.00}{27.00}{25}{23.44}{28.13}{26}
\emline{23.44}{28.13}{27}{25.25}{29.00}{28}
\emline{25.25}{29.00}{29}{27.44}{29.63}{30}
\emline{27.44}{29.63}{31}{30.00}{30.00}{32}
\emline{30.00}{30.00}{33}{32.35}{29.93}{34}
\emline{32.35}{29.93}{35}{34.38}{30.17}{36}
\emline{34.38}{30.17}{37}{36.11}{30.71}{38}
\emline{36.11}{30.71}{39}{37.52}{31.57}{40}
\emline{37.52}{31.57}{41}{38.63}{32.74}{42}
\emline{38.63}{32.74}{43}{39.42}{34.21}{44}
\emline{39.42}{34.21}{45}{39.91}{36.00}{46}
\emline{39.91}{36.00}{47}{40.00}{40.00}{48}
\emline{40.00}{40.00}{49}{39.94}{42.13}{50}
\emline{39.94}{42.13}{51}{40.25}{44.00}{52}
\emline{40.25}{44.00}{53}{40.94}{45.63}{54}
\emline{40.94}{45.63}{55}{42.00}{47.00}{56}
\emline{42.00}{47.00}{57}{43.44}{48.13}{58}
\emline{43.44}{48.13}{59}{45.25}{49.00}{60}
\emline{45.25}{49.00}{61}{47.44}{49.63}{62}
\emline{47.44}{49.63}{63}{50.00}{50.00}{64}
\emline{50.00}{50.00}{65}{52.35}{49.93}{66}
\emline{52.35}{49.93}{67}{54.38}{50.17}{68}
\emline{54.38}{50.17}{69}{56.11}{50.71}{70}
\emline{56.11}{50.71}{71}{57.52}{51.57}{72}
\emline{57.52}{51.57}{73}{58.63}{52.74}{74}
\emline{58.63}{52.74}{75}{59.42}{54.21}{76}
\emline{59.42}{54.21}{77}{59.91}{56.00}{78}
\emline{59.91}{56.00}{79}{60.00}{60.00}{80}
\emline{60.00}{60.00}{81}{59.94}{62.13}{82}
\emline{59.94}{62.13}{83}{60.25}{64.00}{84}
\emline{60.25}{64.00}{85}{60.94}{65.63}{86}
\emline{60.94}{65.63}{87}{62.00}{67.00}{88}
\emline{62.00}{67.00}{89}{63.44}{68.13}{90}
\emline{63.44}{68.13}{91}{65.25}{69.00}{92}
\emline{65.25}{69.00}{93}{67.44}{69.63}{94}
\emline{67.44}{69.63}{95}{70.00}{70.00}{96}
\emline{70.00}{70.00}{97}{72.35}{69.93}{98}
\emline{72.35}{69.93}{99}{74.38}{70.17}{100}
\emline{74.38}{70.17}{101}{76.11}{70.71}{102}
\emline{76.11}{70.71}{103}{77.52}{71.57}{104}
\emline{77.52}{71.57}{105}{78.63}{72.74}{106}
\emline{78.63}{72.74}{107}{79.42}{74.21}{108}
\emline{79.42}{74.21}{109}{79.91}{76.00}{110}
\emline{79.91}{76.00}{111}{80.00}{80.00}{112}
\emline{80.00}{80.00}{113}{79.94}{82.13}{114}
\emline{79.94}{82.13}{115}{80.25}{84.00}{116}
\emline{80.25}{84.00}{117}{80.94}{85.63}{118}
\emline{80.94}{85.63}{119}{82.00}{87.00}{120}
\emline{82.00}{87.00}{121}{83.44}{88.13}{122}
\emline{83.44}{88.13}{123}{85.25}{89.00}{124}
\emline{85.25}{89.00}{125}{87.44}{89.63}{126}
\emline{87.44}{89.63}{127}{90.00}{90.00}{128}
\emline{170.00}{10.00}{129}{170.07}{12.35}{130}
\emline{170.07}{12.35}{131}{169.83}{14.38}{132}
\emline{169.83}{14.38}{133}{169.29}{16.11}{134}
\emline{169.29}{16.11}{135}{168.43}{17.52}{136}
\emline{168.43}{17.52}{137}{167.26}{18.63}{138}
\emline{167.26}{18.63}{139}{165.79}{19.42}{140}
\emline{165.79}{19.42}{141}{164.00}{19.91}{142}
\emline{164.00}{19.91}{143}{160.00}{20.00}{144}
\emline{160.00}{20.00}{145}{157.88}{19.94}{146}
\emline{157.88}{19.94}{147}{156.00}{20.25}{148}
\emline{156.00}{20.25}{149}{154.38}{20.94}{150}
\emline{154.38}{20.94}{151}{153.00}{22.00}{152}
\emline{153.00}{22.00}{153}{151.88}{23.44}{154}
\emline{151.88}{23.44}{155}{151.00}{25.25}{156}
\emline{151.00}{25.25}{157}{150.38}{27.44}{158}
\emline{150.38}{27.44}{159}{150.00}{30.00}{160}
\emline{150.00}{30.00}{161}{150.07}{32.35}{162}
\emline{150.07}{32.35}{163}{149.83}{34.38}{164}
\emline{149.83}{34.38}{165}{149.29}{36.11}{166}
\emline{149.29}{36.11}{167}{148.43}{37.52}{168}
\emline{148.43}{37.52}{169}{147.26}{38.63}{170}
\emline{147.26}{38.63}{171}{145.79}{39.42}{172}
\emline{145.79}{39.42}{173}{144.00}{39.91}{174}
\emline{144.00}{39.91}{175}{140.00}{40.00}{176}
\emline{140.00}{40.00}{177}{137.88}{39.94}{178}
\emline{137.88}{39.94}{179}{136.00}{40.25}{180}
\emline{136.00}{40.25}{181}{134.38}{40.94}{182}
\emline{134.38}{40.94}{183}{133.00}{42.00}{184}
\emline{133.00}{42.00}{185}{131.88}{43.44}{186}
\emline{131.88}{43.44}{187}{131.00}{45.25}{188}
\emline{131.00}{45.25}{189}{130.38}{47.44}{190}
\emline{130.38}{47.44}{191}{130.00}{50.00}{192}
\emline{130.00}{50.00}{193}{130.07}{52.35}{194}
\emline{130.07}{52.35}{195}{129.83}{54.38}{196}
\emline{129.83}{54.38}{197}{129.29}{56.11}{198}
\emline{129.29}{56.11}{199}{128.43}{57.52}{200}
\emline{128.43}{57.52}{201}{127.26}{58.63}{202}
\emline{127.26}{58.63}{203}{125.79}{59.42}{204}
\emline{125.79}{59.42}{205}{124.00}{59.91}{206}
\emline{124.00}{59.91}{207}{120.00}{60.00}{208}
\emline{120.00}{60.00}{209}{117.88}{59.94}{210}
\emline{117.88}{59.94}{211}{116.00}{60.25}{212}
\emline{116.00}{60.25}{213}{114.38}{60.94}{214}
\emline{114.38}{60.94}{215}{113.00}{62.00}{216}
\emline{113.00}{62.00}{217}{111.88}{63.44}{218}
\emline{111.88}{63.44}{219}{111.00}{65.25}{220}
\emline{111.00}{65.25}{221}{110.38}{67.44}{222}
\emline{110.38}{67.44}{223}{110.00}{70.00}{224}
\emline{110.00}{70.00}{225}{110.07}{72.35}{226}
\emline{110.07}{72.35}{227}{109.83}{74.38}{228}
\emline{109.83}{74.38}{229}{109.29}{76.11}{230}
\emline{109.29}{76.11}{231}{108.43}{77.52}{232}
\emline{108.43}{77.52}{233}{107.26}{78.63}{234}
\emline{107.26}{78.63}{235}{105.79}{79.42}{236}
\emline{105.79}{79.42}{237}{104.00}{79.91}{238}
\emline{104.00}{79.91}{239}{100.00}{80.00}{240}
\emline{100.00}{80.00}{241}{97.88}{79.94}{242}
\emline{97.88}{79.94}{243}{96.00}{80.25}{244}
\emline{96.00}{80.25}{245}{94.38}{80.94}{246}
\emline{94.38}{80.94}{247}{93.00}{82.00}{248}
\emline{93.00}{82.00}{249}{91.88}{83.44}{250}
\emline{91.88}{83.44}{251}{91.00}{85.25}{252}
\emline{91.00}{85.25}{253}{90.38}{87.44}{254}
\emline{90.38}{87.44}{255}{90.00}{90.00}{256}
\emline{90.00}{90.00}{257}{87.92}{91.13}{258}
\emline{87.92}{91.13}{259}{86.21}{92.26}{260}
\emline{86.21}{92.26}{261}{84.86}{93.39}{262}
\emline{84.86}{93.39}{263}{83.88}{94.52}{264}
\emline{83.88}{94.52}{265}{83.26}{95.65}{266}
\emline{83.26}{95.65}{267}{83.01}{96.77}{268}
\emline{83.01}{96.77}{269}{83.12}{97.90}{270}
\emline{83.12}{97.90}{271}{83.59}{99.03}{272}
\emline{83.59}{99.03}{273}{84.43}{100.16}{274}
\emline{84.43}{100.16}{275}{85.63}{101.29}{276}
\emline{85.63}{101.29}{277}{87.20}{102.42}{278}
\emline{87.20}{102.42}{279}{90.00}{104.00}{280}
\emline{90.00}{104.00}{281}{92.08}{105.13}{282}
\emline{92.08}{105.13}{283}{93.79}{106.26}{284}
\emline{93.79}{106.26}{285}{95.14}{107.39}{286}
\emline{95.14}{107.39}{287}{96.12}{108.52}{288}
\emline{96.12}{108.52}{289}{96.74}{109.65}{290}
\emline{96.74}{109.65}{291}{96.99}{110.77}{292}
\emline{96.99}{110.77}{293}{96.88}{111.90}{294}
\emline{96.88}{111.90}{295}{96.41}{113.03}{296}
\emline{96.41}{113.03}{297}{95.57}{114.16}{298}
\emline{95.57}{114.16}{299}{94.37}{115.29}{300}
\emline{94.37}{115.29}{301}{92.80}{116.42}{302}
\emline{92.80}{116.42}{303}{90.00}{118.00}{304}
\emline{90.00}{118.00}{305}{87.92}{119.13}{306}
\emline{87.92}{119.13}{307}{86.21}{120.26}{308}
\emline{86.21}{120.26}{309}{84.86}{121.39}{310}
\emline{84.86}{121.39}{311}{83.88}{122.52}{312}
\emline{83.88}{122.52}{313}{83.26}{123.65}{314}
\emline{83.26}{123.65}{315}{83.01}{124.77}{316}
\emline{83.01}{124.77}{317}{83.12}{125.90}{318}
\emline{83.12}{125.90}{319}{83.59}{127.03}{320}
\emline{83.59}{127.03}{321}{84.43}{128.16}{322}
\emline{84.43}{128.16}{323}{85.63}{129.29}{324}
\emline{85.63}{129.29}{325}{87.20}{130.42}{326}
\emline{87.20}{130.42}{327}{90.00}{132.00}{328}
\emline{90.00}{146.00}{329}{87.92}{147.13}{330}
\emline{87.92}{147.13}{331}{86.21}{148.26}{332}
\emline{86.21}{148.26}{333}{84.86}{149.39}{334}
\emline{84.86}{149.39}{335}{83.88}{150.52}{336}
\emline{83.88}{150.52}{337}{83.26}{151.65}{338}
\emline{83.26}{151.65}{339}{83.01}{152.77}{340}
\emline{83.01}{152.77}{341}{83.12}{153.90}{342}
\emline{83.12}{153.90}{343}{83.59}{155.03}{344}
\emline{83.59}{155.03}{345}{84.43}{156.16}{346}
\emline{84.43}{156.16}{347}{85.63}{157.29}{348}
\emline{85.63}{157.29}{349}{87.20}{158.42}{350}
\emline{87.20}{158.42}{351}{90.00}{160.00}{352}
\emline{90.00}{132.00}{353}{92.08}{133.13}{354}
\emline{92.08}{133.13}{355}{93.79}{134.26}{356}
\emline{93.79}{134.26}{357}{95.14}{135.39}{358}
\emline{95.14}{135.39}{359}{96.12}{136.52}{360}
\emline{96.12}{136.52}{361}{96.74}{137.65}{362}
\emline{96.74}{137.65}{363}{96.99}{138.77}{364}
\emline{96.99}{138.77}{365}{96.88}{139.90}{366}
\emline{96.88}{139.90}{367}{96.41}{141.03}{368}
\emline{96.41}{141.03}{369}{95.57}{142.16}{370}
\emline{95.57}{142.16}{371}{94.37}{143.29}{372}
\emline{94.37}{143.29}{373}{92.80}{144.42}{374}
\emline{92.80}{144.42}{375}{90.00}{146.00}{376}
\emline{90.00}{160.00}{377}{92.08}{161.13}{378}
\emline{92.08}{161.13}{379}{93.79}{162.26}{380}
\emline{93.79}{162.26}{381}{95.14}{163.39}{382}
\emline{95.14}{163.39}{383}{96.12}{164.52}{384}
\emline{96.12}{164.52}{385}{96.74}{165.65}{386}
\emline{96.74}{165.65}{387}{96.99}{166.77}{388}
\emline{96.99}{166.77}{389}{96.88}{167.90}{390}
\emline{96.88}{167.90}{391}{96.41}{169.03}{392}
\emline{96.41}{169.03}{393}{95.57}{170.16}{394}
\emline{95.57}{170.16}{395}{94.37}{171.29}{396}
\emline{94.37}{171.29}{397}{92.80}{172.42}{398}
\emline{92.80}{172.42}{399}{90.00}{174.00}{400}
\put(90.00,90.00){\circle*{2.00}}
\put(6.00,0.00){\makebox(0,0)[ct]{$W^+,\mu$}}
\put(170.00,0.00){\makebox(0,0)[ct]{$W^-,\nu$}}
\put(120.00,170.00){\makebox(0,0)[lc]{$\gamma (Z), \rho$}}
\put(25.00,55.00){\makebox(0,0)[cc]{$k$}}
\put(155.00,55.00){\makebox(0,0)[cc]{$p$}}
\put(60.00,120.00){\makebox(0,0)[cc]{$q$}}
\put(55.00,53.00){\vector(-1,-1){0.2}}
\put(135.00,42.67){\vector(1,-1){0.2}}
\put(83.33,130.00){\vector(0,1){0.2}}
\end{picture}
\emp
$i e (i g\cos{\theta_W}) V_{\nu\mu\rho}(k,p,q)$\hfill
\bmp{3.5cm}
\special{em:linewidth 0.3pt}
\unitlength 0.15mm
\linethickness{0.3pt}
\begin{picture}(180.00,180.00)
\emline{10.00}{10.00}{1}{12.35}{9.93}{2}
\emline{12.35}{9.93}{3}{14.38}{10.17}{4}
\emline{14.38}{10.17}{5}{16.11}{10.71}{6}
\emline{16.11}{10.71}{7}{17.52}{11.57}{8}
\emline{17.52}{11.57}{9}{18.63}{12.74}{10}
\emline{18.63}{12.74}{11}{19.42}{14.21}{12}
\emline{19.42}{14.21}{13}{19.91}{16.00}{14}
\emline{19.91}{16.00}{15}{20.00}{20.00}{16}
\emline{20.00}{20.00}{17}{19.94}{22.13}{18}
\emline{19.94}{22.13}{19}{20.25}{24.00}{20}
\emline{20.25}{24.00}{21}{20.94}{25.63}{22}
\emline{20.94}{25.63}{23}{22.00}{27.00}{24}
\emline{22.00}{27.00}{25}{23.44}{28.13}{26}
\emline{23.44}{28.13}{27}{25.25}{29.00}{28}
\emline{25.25}{29.00}{29}{27.44}{29.63}{30}
\emline{27.44}{29.63}{31}{30.00}{30.00}{32}
\emline{30.00}{30.00}{33}{32.35}{29.93}{34}
\emline{32.35}{29.93}{35}{34.38}{30.17}{36}
\emline{34.38}{30.17}{37}{36.11}{30.71}{38}
\emline{36.11}{30.71}{39}{37.52}{31.57}{40}
\emline{37.52}{31.57}{41}{38.63}{32.74}{42}
\emline{38.63}{32.74}{43}{39.42}{34.21}{44}
\emline{39.42}{34.21}{45}{39.91}{36.00}{46}
\emline{39.91}{36.00}{47}{40.00}{40.00}{48}
\emline{40.00}{40.00}{49}{39.94}{42.13}{50}
\emline{39.94}{42.13}{51}{40.25}{44.00}{52}
\emline{40.25}{44.00}{53}{40.94}{45.63}{54}
\emline{40.94}{45.63}{55}{42.00}{47.00}{56}
\emline{42.00}{47.00}{57}{43.44}{48.13}{58}
\emline{43.44}{48.13}{59}{45.25}{49.00}{60}
\emline{45.25}{49.00}{61}{47.44}{49.63}{62}
\emline{47.44}{49.63}{63}{50.00}{50.00}{64}
\emline{50.00}{50.00}{65}{52.35}{49.93}{66}
\emline{52.35}{49.93}{67}{54.38}{50.17}{68}
\emline{54.38}{50.17}{69}{56.11}{50.71}{70}
\emline{56.11}{50.71}{71}{57.52}{51.57}{72}
\emline{57.52}{51.57}{73}{58.63}{52.74}{74}
\emline{58.63}{52.74}{75}{59.42}{54.21}{76}
\emline{59.42}{54.21}{77}{59.91}{56.00}{78}
\emline{59.91}{56.00}{79}{60.00}{60.00}{80}
\emline{60.00}{60.00}{81}{59.94}{62.13}{82}
\emline{59.94}{62.13}{83}{60.25}{64.00}{84}
\emline{60.25}{64.00}{85}{60.94}{65.63}{86}
\emline{60.94}{65.63}{87}{62.00}{67.00}{88}
\emline{62.00}{67.00}{89}{63.44}{68.13}{90}
\emline{63.44}{68.13}{91}{65.25}{69.00}{92}
\emline{65.25}{69.00}{93}{67.44}{69.63}{94}
\emline{67.44}{69.63}{95}{70.00}{70.00}{96}
\emline{70.00}{70.00}{97}{72.35}{69.93}{98}
\emline{72.35}{69.93}{99}{74.38}{70.17}{100}
\emline{74.38}{70.17}{101}{76.11}{70.71}{102}
\emline{76.11}{70.71}{103}{77.52}{71.57}{104}
\emline{77.52}{71.57}{105}{78.63}{72.74}{106}
\emline{78.63}{72.74}{107}{79.42}{74.21}{108}
\emline{79.42}{74.21}{109}{79.91}{76.00}{110}
\emline{79.91}{76.00}{111}{80.00}{80.00}{112}
\emline{80.00}{80.00}{113}{79.94}{82.13}{114}
\emline{79.94}{82.13}{115}{80.25}{84.00}{116}
\emline{80.25}{84.00}{117}{80.94}{85.63}{118}
\emline{80.94}{85.63}{119}{82.00}{87.00}{120}
\emline{82.00}{87.00}{121}{83.44}{88.13}{122}
\emline{83.44}{88.13}{123}{85.25}{89.00}{124}
\emline{85.25}{89.00}{125}{87.44}{89.63}{126}
\emline{87.44}{89.63}{127}{90.00}{90.00}{128}
\emline{170.00}{10.00}{129}{170.07}{12.35}{130}
\emline{170.07}{12.35}{131}{169.83}{14.38}{132}
\emline{169.83}{14.38}{133}{169.29}{16.11}{134}
\emline{169.29}{16.11}{135}{168.43}{17.52}{136}
\emline{168.43}{17.52}{137}{167.26}{18.63}{138}
\emline{167.26}{18.63}{139}{165.79}{19.42}{140}
\emline{165.79}{19.42}{141}{164.00}{19.91}{142}
\emline{164.00}{19.91}{143}{160.00}{20.00}{144}
\emline{160.00}{20.00}{145}{157.88}{19.94}{146}
\emline{157.88}{19.94}{147}{156.00}{20.25}{148}
\emline{156.00}{20.25}{149}{154.38}{20.94}{150}
\emline{154.38}{20.94}{151}{153.00}{22.00}{152}
\emline{153.00}{22.00}{153}{151.88}{23.44}{154}
\emline{151.88}{23.44}{155}{151.00}{25.25}{156}
\emline{151.00}{25.25}{157}{150.38}{27.44}{158}
\emline{150.38}{27.44}{159}{150.00}{30.00}{160}
\emline{150.00}{30.00}{161}{150.07}{32.35}{162}
\emline{150.07}{32.35}{163}{149.83}{34.38}{164}
\emline{149.83}{34.38}{165}{149.29}{36.11}{166}
\emline{149.29}{36.11}{167}{148.43}{37.52}{168}
\emline{148.43}{37.52}{169}{147.26}{38.63}{170}
\emline{147.26}{38.63}{171}{145.79}{39.42}{172}
\emline{145.79}{39.42}{173}{144.00}{39.91}{174}
\emline{144.00}{39.91}{175}{140.00}{40.00}{176}
\emline{140.00}{40.00}{177}{137.88}{39.94}{178}
\emline{137.88}{39.94}{179}{136.00}{40.25}{180}
\emline{136.00}{40.25}{181}{134.38}{40.94}{182}
\emline{134.38}{40.94}{183}{133.00}{42.00}{184}
\emline{133.00}{42.00}{185}{131.88}{43.44}{186}
\emline{131.88}{43.44}{187}{131.00}{45.25}{188}
\emline{131.00}{45.25}{189}{130.38}{47.44}{190}
\emline{130.38}{47.44}{191}{130.00}{50.00}{192}
\emline{130.00}{50.00}{193}{130.07}{52.35}{194}
\emline{130.07}{52.35}{195}{129.83}{54.38}{196}
\emline{129.83}{54.38}{197}{129.29}{56.11}{198}
\emline{129.29}{56.11}{199}{128.43}{57.52}{200}
\emline{128.43}{57.52}{201}{127.26}{58.63}{202}
\emline{127.26}{58.63}{203}{125.79}{59.42}{204}
\emline{125.79}{59.42}{205}{124.00}{59.91}{206}
\emline{124.00}{59.91}{207}{120.00}{60.00}{208}
\emline{120.00}{60.00}{209}{117.88}{59.94}{210}
\emline{117.88}{59.94}{211}{116.00}{60.25}{212}
\emline{116.00}{60.25}{213}{114.38}{60.94}{214}
\emline{114.38}{60.94}{215}{113.00}{62.00}{216}
\emline{113.00}{62.00}{217}{111.88}{63.44}{218}
\emline{111.88}{63.44}{219}{111.00}{65.25}{220}
\emline{111.00}{65.25}{221}{110.38}{67.44}{222}
\emline{110.38}{67.44}{223}{110.00}{70.00}{224}
\emline{110.00}{70.00}{225}{110.07}{72.35}{226}
\emline{110.07}{72.35}{227}{109.83}{74.38}{228}
\emline{109.83}{74.38}{229}{109.29}{76.11}{230}
\emline{109.29}{76.11}{231}{108.43}{77.52}{232}
\emline{108.43}{77.52}{233}{107.26}{78.63}{234}
\emline{107.26}{78.63}{235}{105.79}{79.42}{236}
\emline{105.79}{79.42}{237}{104.00}{79.91}{238}
\emline{104.00}{79.91}{239}{100.00}{80.00}{240}
\emline{100.00}{80.00}{241}{97.88}{79.94}{242}
\emline{97.88}{79.94}{243}{96.00}{80.25}{244}
\emline{96.00}{80.25}{245}{94.38}{80.94}{246}
\emline{94.38}{80.94}{247}{93.00}{82.00}{248}
\emline{93.00}{82.00}{249}{91.88}{83.44}{250}
\emline{91.88}{83.44}{251}{91.00}{85.25}{252}
\emline{91.00}{85.25}{253}{90.38}{87.44}{254}
\emline{90.38}{87.44}{255}{90.00}{90.00}{256}
\emline{10.00}{170.00}{257}{9.93}{167.65}{258}
\emline{9.93}{167.65}{259}{10.17}{165.62}{260}
\emline{10.17}{165.62}{261}{10.71}{163.89}{262}
\emline{10.71}{163.89}{263}{11.57}{162.48}{264}
\emline{11.57}{162.48}{265}{12.74}{161.37}{266}
\emline{12.74}{161.37}{267}{14.21}{160.58}{268}
\emline{14.21}{160.58}{269}{16.00}{160.09}{270}
\emline{16.00}{160.09}{271}{20.00}{160.00}{272}
\emline{20.00}{160.00}{273}{22.13}{160.06}{274}
\emline{22.13}{160.06}{275}{24.00}{159.75}{276}
\emline{24.00}{159.75}{277}{25.63}{159.06}{278}
\emline{25.63}{159.06}{279}{27.00}{158.00}{280}
\emline{27.00}{158.00}{281}{28.13}{156.56}{282}
\emline{28.13}{156.56}{283}{29.00}{154.75}{284}
\emline{29.00}{154.75}{285}{29.63}{152.56}{286}
\emline{29.63}{152.56}{287}{30.00}{150.00}{288}
\emline{30.00}{150.00}{289}{29.93}{147.65}{290}
\emline{29.93}{147.65}{291}{30.17}{145.62}{292}
\emline{30.17}{145.62}{293}{30.71}{143.89}{294}
\emline{30.71}{143.89}{295}{31.57}{142.48}{296}
\emline{31.57}{142.48}{297}{32.74}{141.37}{298}
\emline{32.74}{141.37}{299}{34.21}{140.58}{300}
\emline{34.21}{140.58}{301}{36.00}{140.09}{302}
\emline{36.00}{140.09}{303}{40.00}{140.00}{304}
\emline{40.00}{140.00}{305}{42.13}{140.06}{306}
\emline{42.13}{140.06}{307}{44.00}{139.75}{308}
\emline{44.00}{139.75}{309}{45.63}{139.06}{310}
\emline{45.63}{139.06}{311}{47.00}{138.00}{312}
\emline{47.00}{138.00}{313}{48.13}{136.56}{314}
\emline{48.13}{136.56}{315}{49.00}{134.75}{316}
\emline{49.00}{134.75}{317}{49.63}{132.56}{318}
\emline{49.63}{132.56}{319}{50.00}{130.00}{320}
\emline{50.00}{130.00}{321}{49.93}{127.65}{322}
\emline{49.93}{127.65}{323}{50.17}{125.62}{324}
\emline{50.17}{125.62}{325}{50.71}{123.89}{326}
\emline{50.71}{123.89}{327}{51.57}{122.48}{328}
\emline{51.57}{122.48}{329}{52.74}{121.37}{330}
\emline{52.74}{121.37}{331}{54.21}{120.58}{332}
\emline{54.21}{120.58}{333}{56.00}{120.09}{334}
\emline{56.00}{120.09}{335}{60.00}{120.00}{336}
\emline{60.00}{120.00}{337}{62.13}{120.06}{338}
\emline{62.13}{120.06}{339}{64.00}{119.75}{340}
\emline{64.00}{119.75}{341}{65.63}{119.06}{342}
\emline{65.63}{119.06}{343}{67.00}{118.00}{344}
\emline{67.00}{118.00}{345}{68.13}{116.56}{346}
\emline{68.13}{116.56}{347}{69.00}{114.75}{348}
\emline{69.00}{114.75}{349}{69.63}{112.56}{350}
\emline{69.63}{112.56}{351}{70.00}{110.00}{352}
\emline{70.00}{110.00}{353}{69.93}{107.65}{354}
\emline{69.93}{107.65}{355}{70.17}{105.62}{356}
\emline{70.17}{105.62}{357}{70.71}{103.89}{358}
\emline{70.71}{103.89}{359}{71.57}{102.48}{360}
\emline{71.57}{102.48}{361}{72.74}{101.37}{362}
\emline{72.74}{101.37}{363}{74.21}{100.58}{364}
\emline{74.21}{100.58}{365}{76.00}{100.09}{366}
\emline{76.00}{100.09}{367}{80.00}{100.00}{368}
\emline{80.00}{100.00}{369}{82.13}{100.06}{370}
\emline{82.13}{100.06}{371}{84.00}{99.75}{372}
\emline{84.00}{99.75}{373}{85.63}{99.06}{374}
\emline{85.63}{99.06}{375}{87.00}{98.00}{376}
\emline{87.00}{98.00}{377}{88.13}{96.56}{378}
\emline{88.13}{96.56}{379}{89.00}{94.75}{380}
\emline{89.00}{94.75}{381}{89.63}{92.56}{382}
\emline{89.63}{92.56}{383}{90.00}{90.00}{384}
\emline{170.00}{170.00}{385}{167.65}{170.07}{386}
\emline{167.65}{170.07}{387}{165.62}{169.83}{388}
\emline{165.62}{169.83}{389}{163.89}{169.29}{390}
\emline{163.89}{169.29}{391}{162.48}{168.43}{392}
\emline{162.48}{168.43}{393}{161.37}{167.26}{394}
\emline{161.37}{167.26}{395}{160.58}{165.79}{396}
\emline{160.58}{165.79}{397}{160.09}{164.00}{398}
\emline{160.09}{164.00}{399}{160.00}{160.00}{400}
\emline{160.00}{160.00}{401}{160.06}{157.88}{402}
\emline{160.06}{157.88}{403}{159.75}{156.00}{404}
\emline{159.75}{156.00}{405}{159.06}{154.38}{406}
\emline{159.06}{154.38}{407}{158.00}{153.00}{408}
\emline{158.00}{153.00}{409}{156.56}{151.88}{410}
\emline{156.56}{151.88}{411}{154.75}{151.00}{412}
\emline{154.75}{151.00}{413}{152.56}{150.38}{414}
\emline{152.56}{150.38}{415}{150.00}{150.00}{416}
\emline{150.00}{150.00}{417}{147.65}{150.07}{418}
\emline{147.65}{150.07}{419}{145.62}{149.83}{420}
\emline{145.62}{149.83}{421}{143.89}{149.29}{422}
\emline{143.89}{149.29}{423}{142.48}{148.43}{424}
\emline{142.48}{148.43}{425}{141.37}{147.26}{426}
\emline{141.37}{147.26}{427}{140.58}{145.79}{428}
\emline{140.58}{145.79}{429}{140.09}{144.00}{430}
\emline{140.09}{144.00}{431}{140.00}{140.00}{432}
\emline{140.00}{140.00}{433}{140.06}{137.88}{434}
\emline{140.06}{137.88}{435}{139.75}{136.00}{436}
\emline{139.75}{136.00}{437}{139.06}{134.38}{438}
\emline{139.06}{134.38}{439}{138.00}{133.00}{440}
\emline{138.00}{133.00}{441}{136.56}{131.88}{442}
\emline{136.56}{131.88}{443}{134.75}{131.00}{444}
\emline{134.75}{131.00}{445}{132.56}{130.38}{446}
\emline{132.56}{130.38}{447}{130.00}{130.00}{448}
\emline{130.00}{130.00}{449}{127.65}{130.07}{450}
\emline{127.65}{130.07}{451}{125.62}{129.83}{452}
\emline{125.62}{129.83}{453}{123.89}{129.29}{454}
\emline{123.89}{129.29}{455}{122.48}{128.43}{456}
\emline{122.48}{128.43}{457}{121.37}{127.26}{458}
\emline{121.37}{127.26}{459}{120.58}{125.79}{460}
\emline{120.58}{125.79}{461}{120.09}{124.00}{462}
\emline{120.09}{124.00}{463}{120.00}{120.00}{464}
\emline{120.00}{120.00}{465}{120.06}{117.88}{466}
\emline{120.06}{117.88}{467}{119.75}{116.00}{468}
\emline{119.75}{116.00}{469}{119.06}{114.38}{470}
\emline{119.06}{114.38}{471}{118.00}{113.00}{472}
\emline{118.00}{113.00}{473}{116.56}{111.88}{474}
\emline{116.56}{111.88}{475}{114.75}{111.00}{476}
\emline{114.75}{111.00}{477}{112.56}{110.38}{478}
\emline{112.56}{110.38}{479}{110.00}{110.00}{480}
\emline{110.00}{110.00}{481}{107.65}{110.07}{482}
\emline{107.65}{110.07}{483}{105.62}{109.83}{484}
\emline{105.62}{109.83}{485}{103.89}{109.29}{486}
\emline{103.89}{109.29}{487}{102.48}{108.43}{488}
\emline{102.48}{108.43}{489}{101.37}{107.26}{490}
\emline{101.37}{107.26}{491}{100.58}{105.79}{492}
\emline{100.58}{105.79}{493}{100.09}{104.00}{494}
\emline{100.09}{104.00}{495}{100.00}{100.00}{496}
\emline{100.00}{100.00}{497}{100.06}{97.88}{498}
\emline{100.06}{97.88}{499}{99.75}{96.00}{500}
\emline{99.75}{96.00}{501}{99.06}{94.38}{502}
\emline{99.06}{94.38}{503}{98.00}{93.00}{504}
\emline{98.00}{93.00}{505}{96.56}{91.88}{506}
\emline{96.56}{91.88}{507}{94.75}{91.00}{508}
\emline{94.75}{91.00}{509}{92.56}{90.38}{510}
\emline{92.56}{90.38}{511}{90.00}{90.00}{512}
\put(90.00,90.00){\circle*{3.00}}
\put(7.00,0.00){\makebox(0,0)[ct]{$W^+,\mu$}}
\put(180.00,0.00){\makebox(0,0)[ct]{$W^-,\lambda$}}
\put(7.00,180.00){\makebox(0,0)[cc]{$W^-,\sigma$}}
\put(180.00,180.00){\makebox(0,0)[cc]{$W^+,\nu$}}
\put(55.00,53.00){\vector(-1,-1){0.2}}
\put(135.00,44.00){\vector(1,-1){0.2}}
\put(125.00,127.00){\vector(1,1){0.2}}
\put(45.00,136.00){\vector(-1,1){0.2}}
\end{picture}
\emp
$i g^2 V_{\mu\nu\lambda\sigma}$\\[10mm]
\bmp{3.5cm}
\special{em:linewidth 0.3pt}
\unitlength 0.15mm
\linethickness{0.3pt}
\begin{picture}(180.00,180.00)
\emline{10.00}{10.00}{1}{12.35}{9.93}{2}
\emline{12.35}{9.93}{3}{14.38}{10.17}{4}
\emline{14.38}{10.17}{5}{16.11}{10.71}{6}
\emline{16.11}{10.71}{7}{17.52}{11.57}{8}
\emline{17.52}{11.57}{9}{18.63}{12.74}{10}
\emline{18.63}{12.74}{11}{19.42}{14.21}{12}
\emline{19.42}{14.21}{13}{19.91}{16.00}{14}
\emline{19.91}{16.00}{15}{20.00}{20.00}{16}
\emline{20.00}{20.00}{17}{19.94}{22.13}{18}
\emline{19.94}{22.13}{19}{20.25}{24.00}{20}
\emline{20.25}{24.00}{21}{20.94}{25.63}{22}
\emline{20.94}{25.63}{23}{22.00}{27.00}{24}
\emline{22.00}{27.00}{25}{23.44}{28.13}{26}
\emline{23.44}{28.13}{27}{25.25}{29.00}{28}
\emline{25.25}{29.00}{29}{27.44}{29.63}{30}
\emline{27.44}{29.63}{31}{30.00}{30.00}{32}
\emline{30.00}{30.00}{33}{32.35}{29.93}{34}
\emline{32.35}{29.93}{35}{34.38}{30.17}{36}
\emline{34.38}{30.17}{37}{36.11}{30.71}{38}
\emline{36.11}{30.71}{39}{37.52}{31.57}{40}
\emline{37.52}{31.57}{41}{38.63}{32.74}{42}
\emline{38.63}{32.74}{43}{39.42}{34.21}{44}
\emline{39.42}{34.21}{45}{39.91}{36.00}{46}
\emline{39.91}{36.00}{47}{40.00}{40.00}{48}
\emline{40.00}{40.00}{49}{39.94}{42.13}{50}
\emline{39.94}{42.13}{51}{40.25}{44.00}{52}
\emline{40.25}{44.00}{53}{40.94}{45.63}{54}
\emline{40.94}{45.63}{55}{42.00}{47.00}{56}
\emline{42.00}{47.00}{57}{43.44}{48.13}{58}
\emline{43.44}{48.13}{59}{45.25}{49.00}{60}
\emline{45.25}{49.00}{61}{47.44}{49.63}{62}
\emline{47.44}{49.63}{63}{50.00}{50.00}{64}
\emline{50.00}{50.00}{65}{52.35}{49.93}{66}
\emline{52.35}{49.93}{67}{54.38}{50.17}{68}
\emline{54.38}{50.17}{69}{56.11}{50.71}{70}
\emline{56.11}{50.71}{71}{57.52}{51.57}{72}
\emline{57.52}{51.57}{73}{58.63}{52.74}{74}
\emline{58.63}{52.74}{75}{59.42}{54.21}{76}
\emline{59.42}{54.21}{77}{59.91}{56.00}{78}
\emline{59.91}{56.00}{79}{60.00}{60.00}{80}
\emline{60.00}{60.00}{81}{59.94}{62.13}{82}
\emline{59.94}{62.13}{83}{60.25}{64.00}{84}
\emline{60.25}{64.00}{85}{60.94}{65.63}{86}
\emline{60.94}{65.63}{87}{62.00}{67.00}{88}
\emline{62.00}{67.00}{89}{63.44}{68.13}{90}
\emline{63.44}{68.13}{91}{65.25}{69.00}{92}
\emline{65.25}{69.00}{93}{67.44}{69.63}{94}
\emline{67.44}{69.63}{95}{70.00}{70.00}{96}
\emline{70.00}{70.00}{97}{72.35}{69.93}{98}
\emline{72.35}{69.93}{99}{74.38}{70.17}{100}
\emline{74.38}{70.17}{101}{76.11}{70.71}{102}
\emline{76.11}{70.71}{103}{77.52}{71.57}{104}
\emline{77.52}{71.57}{105}{78.63}{72.74}{106}
\emline{78.63}{72.74}{107}{79.42}{74.21}{108}
\emline{79.42}{74.21}{109}{79.91}{76.00}{110}
\emline{79.91}{76.00}{111}{80.00}{80.00}{112}
\emline{80.00}{80.00}{113}{79.94}{82.13}{114}
\emline{79.94}{82.13}{115}{80.25}{84.00}{116}
\emline{80.25}{84.00}{117}{80.94}{85.63}{118}
\emline{80.94}{85.63}{119}{82.00}{87.00}{120}
\emline{82.00}{87.00}{121}{83.44}{88.13}{122}
\emline{83.44}{88.13}{123}{85.25}{89.00}{124}
\emline{85.25}{89.00}{125}{87.44}{89.63}{126}
\emline{87.44}{89.63}{127}{90.00}{90.00}{128}
\emline{170.00}{10.00}{129}{170.07}{12.35}{130}
\emline{170.07}{12.35}{131}{169.83}{14.38}{132}
\emline{169.83}{14.38}{133}{169.29}{16.11}{134}
\emline{169.29}{16.11}{135}{168.43}{17.52}{136}
\emline{168.43}{17.52}{137}{167.26}{18.63}{138}
\emline{167.26}{18.63}{139}{165.79}{19.42}{140}
\emline{165.79}{19.42}{141}{164.00}{19.91}{142}
\emline{164.00}{19.91}{143}{160.00}{20.00}{144}
\emline{160.00}{20.00}{145}{157.88}{19.94}{146}
\emline{157.88}{19.94}{147}{156.00}{20.25}{148}
\emline{156.00}{20.25}{149}{154.38}{20.94}{150}
\emline{154.38}{20.94}{151}{153.00}{22.00}{152}
\emline{153.00}{22.00}{153}{151.88}{23.44}{154}
\emline{151.88}{23.44}{155}{151.00}{25.25}{156}
\emline{151.00}{25.25}{157}{150.38}{27.44}{158}
\emline{150.38}{27.44}{159}{150.00}{30.00}{160}
\emline{150.00}{30.00}{161}{150.07}{32.35}{162}
\emline{150.07}{32.35}{163}{149.83}{34.38}{164}
\emline{149.83}{34.38}{165}{149.29}{36.11}{166}
\emline{149.29}{36.11}{167}{148.43}{37.52}{168}
\emline{148.43}{37.52}{169}{147.26}{38.63}{170}
\emline{147.26}{38.63}{171}{145.79}{39.42}{172}
\emline{145.79}{39.42}{173}{144.00}{39.91}{174}
\emline{144.00}{39.91}{175}{140.00}{40.00}{176}
\emline{140.00}{40.00}{177}{137.88}{39.94}{178}
\emline{137.88}{39.94}{179}{136.00}{40.25}{180}
\emline{136.00}{40.25}{181}{134.38}{40.94}{182}
\emline{134.38}{40.94}{183}{133.00}{42.00}{184}
\emline{133.00}{42.00}{185}{131.88}{43.44}{186}
\emline{131.88}{43.44}{187}{131.00}{45.25}{188}
\emline{131.00}{45.25}{189}{130.38}{47.44}{190}
\emline{130.38}{47.44}{191}{130.00}{50.00}{192}
\emline{130.00}{50.00}{193}{130.07}{52.35}{194}
\emline{130.07}{52.35}{195}{129.83}{54.38}{196}
\emline{129.83}{54.38}{197}{129.29}{56.11}{198}
\emline{129.29}{56.11}{199}{128.43}{57.52}{200}
\emline{128.43}{57.52}{201}{127.26}{58.63}{202}
\emline{127.26}{58.63}{203}{125.79}{59.42}{204}
\emline{125.79}{59.42}{205}{124.00}{59.91}{206}
\emline{124.00}{59.91}{207}{120.00}{60.00}{208}
\emline{120.00}{60.00}{209}{117.88}{59.94}{210}
\emline{117.88}{59.94}{211}{116.00}{60.25}{212}
\emline{116.00}{60.25}{213}{114.38}{60.94}{214}
\emline{114.38}{60.94}{215}{113.00}{62.00}{216}
\emline{113.00}{62.00}{217}{111.88}{63.44}{218}
\emline{111.88}{63.44}{219}{111.00}{65.25}{220}
\emline{111.00}{65.25}{221}{110.38}{67.44}{222}
\emline{110.38}{67.44}{223}{110.00}{70.00}{224}
\emline{110.00}{70.00}{225}{110.07}{72.35}{226}
\emline{110.07}{72.35}{227}{109.83}{74.38}{228}
\emline{109.83}{74.38}{229}{109.29}{76.11}{230}
\emline{109.29}{76.11}{231}{108.43}{77.52}{232}
\emline{108.43}{77.52}{233}{107.26}{78.63}{234}
\emline{107.26}{78.63}{235}{105.79}{79.42}{236}
\emline{105.79}{79.42}{237}{104.00}{79.91}{238}
\emline{104.00}{79.91}{239}{100.00}{80.00}{240}
\emline{100.00}{80.00}{241}{97.88}{79.94}{242}
\emline{97.88}{79.94}{243}{96.00}{80.25}{244}
\emline{96.00}{80.25}{245}{94.38}{80.94}{246}
\emline{94.38}{80.94}{247}{93.00}{82.00}{248}
\emline{93.00}{82.00}{249}{91.88}{83.44}{250}
\emline{91.88}{83.44}{251}{91.00}{85.25}{252}
\emline{91.00}{85.25}{253}{90.38}{87.44}{254}
\emline{90.38}{87.44}{255}{90.00}{90.00}{256}
\emline{10.00}{170.00}{257}{9.93}{167.65}{258}
\emline{9.93}{167.65}{259}{10.17}{165.62}{260}
\emline{10.17}{165.62}{261}{10.71}{163.89}{262}
\emline{10.71}{163.89}{263}{11.57}{162.48}{264}
\emline{11.57}{162.48}{265}{12.74}{161.37}{266}
\emline{12.74}{161.37}{267}{14.21}{160.58}{268}
\emline{14.21}{160.58}{269}{16.00}{160.09}{270}
\emline{16.00}{160.09}{271}{20.00}{160.00}{272}
\emline{20.00}{160.00}{273}{22.13}{160.06}{274}
\emline{22.13}{160.06}{275}{24.00}{159.75}{276}
\emline{24.00}{159.75}{277}{25.63}{159.06}{278}
\emline{25.63}{159.06}{279}{27.00}{158.00}{280}
\emline{27.00}{158.00}{281}{28.13}{156.56}{282}
\emline{28.13}{156.56}{283}{29.00}{154.75}{284}
\emline{29.00}{154.75}{285}{29.63}{152.56}{286}
\emline{29.63}{152.56}{287}{30.00}{150.00}{288}
\emline{30.00}{150.00}{289}{29.93}{147.65}{290}
\emline{29.93}{147.65}{291}{30.17}{145.62}{292}
\emline{30.17}{145.62}{293}{30.71}{143.89}{294}
\emline{30.71}{143.89}{295}{31.57}{142.48}{296}
\emline{31.57}{142.48}{297}{32.74}{141.37}{298}
\emline{32.74}{141.37}{299}{34.21}{140.58}{300}
\emline{34.21}{140.58}{301}{36.00}{140.09}{302}
\emline{36.00}{140.09}{303}{40.00}{140.00}{304}
\emline{40.00}{140.00}{305}{42.13}{140.06}{306}
\emline{42.13}{140.06}{307}{44.00}{139.75}{308}
\emline{44.00}{139.75}{309}{45.63}{139.06}{310}
\emline{45.63}{139.06}{311}{47.00}{138.00}{312}
\emline{47.00}{138.00}{313}{48.13}{136.56}{314}
\emline{48.13}{136.56}{315}{49.00}{134.75}{316}
\emline{49.00}{134.75}{317}{49.63}{132.56}{318}
\emline{49.63}{132.56}{319}{50.00}{130.00}{320}
\emline{50.00}{130.00}{321}{49.93}{127.65}{322}
\emline{49.93}{127.65}{323}{50.17}{125.62}{324}
\emline{50.17}{125.62}{325}{50.71}{123.89}{326}
\emline{50.71}{123.89}{327}{51.57}{122.48}{328}
\emline{51.57}{122.48}{329}{52.74}{121.37}{330}
\emline{52.74}{121.37}{331}{54.21}{120.58}{332}
\emline{54.21}{120.58}{333}{56.00}{120.09}{334}
\emline{56.00}{120.09}{335}{60.00}{120.00}{336}
\emline{60.00}{120.00}{337}{62.13}{120.06}{338}
\emline{62.13}{120.06}{339}{64.00}{119.75}{340}
\emline{64.00}{119.75}{341}{65.63}{119.06}{342}
\emline{65.63}{119.06}{343}{67.00}{118.00}{344}
\emline{67.00}{118.00}{345}{68.13}{116.56}{346}
\emline{68.13}{116.56}{347}{69.00}{114.75}{348}
\emline{69.00}{114.75}{349}{69.63}{112.56}{350}
\emline{69.63}{112.56}{351}{70.00}{110.00}{352}
\emline{70.00}{110.00}{353}{69.93}{107.65}{354}
\emline{69.93}{107.65}{355}{70.17}{105.62}{356}
\emline{70.17}{105.62}{357}{70.71}{103.89}{358}
\emline{70.71}{103.89}{359}{71.57}{102.48}{360}
\emline{71.57}{102.48}{361}{72.74}{101.37}{362}
\emline{72.74}{101.37}{363}{74.21}{100.58}{364}
\emline{74.21}{100.58}{365}{76.00}{100.09}{366}
\emline{76.00}{100.09}{367}{80.00}{100.00}{368}
\emline{80.00}{100.00}{369}{82.13}{100.06}{370}
\emline{82.13}{100.06}{371}{84.00}{99.75}{372}
\emline{84.00}{99.75}{373}{85.63}{99.06}{374}
\emline{85.63}{99.06}{375}{87.00}{98.00}{376}
\emline{87.00}{98.00}{377}{88.13}{96.56}{378}
\emline{88.13}{96.56}{379}{89.00}{94.75}{380}
\emline{89.00}{94.75}{381}{89.63}{92.56}{382}
\emline{89.63}{92.56}{383}{90.00}{90.00}{384}
\emline{170.00}{170.00}{385}{167.65}{170.07}{386}
\emline{167.65}{170.07}{387}{165.62}{169.83}{388}
\emline{165.62}{169.83}{389}{163.89}{169.29}{390}
\emline{163.89}{169.29}{391}{162.48}{168.43}{392}
\emline{162.48}{168.43}{393}{161.37}{167.26}{394}
\emline{161.37}{167.26}{395}{160.58}{165.79}{396}
\emline{160.58}{165.79}{397}{160.09}{164.00}{398}
\emline{160.09}{164.00}{399}{160.00}{160.00}{400}
\emline{160.00}{160.00}{401}{160.06}{157.88}{402}
\emline{160.06}{157.88}{403}{159.75}{156.00}{404}
\emline{159.75}{156.00}{405}{159.06}{154.38}{406}
\emline{159.06}{154.38}{407}{158.00}{153.00}{408}
\emline{158.00}{153.00}{409}{156.56}{151.88}{410}
\emline{156.56}{151.88}{411}{154.75}{151.00}{412}
\emline{154.75}{151.00}{413}{152.56}{150.38}{414}
\emline{152.56}{150.38}{415}{150.00}{150.00}{416}
\emline{150.00}{150.00}{417}{147.65}{150.07}{418}
\emline{147.65}{150.07}{419}{145.62}{149.83}{420}
\emline{145.62}{149.83}{421}{143.89}{149.29}{422}
\emline{143.89}{149.29}{423}{142.48}{148.43}{424}
\emline{142.48}{148.43}{425}{141.37}{147.26}{426}
\emline{141.37}{147.26}{427}{140.58}{145.79}{428}
\emline{140.58}{145.79}{429}{140.09}{144.00}{430}
\emline{140.09}{144.00}{431}{140.00}{140.00}{432}
\emline{140.00}{140.00}{433}{140.06}{137.88}{434}
\emline{140.06}{137.88}{435}{139.75}{136.00}{436}
\emline{139.75}{136.00}{437}{139.06}{134.38}{438}
\emline{139.06}{134.38}{439}{138.00}{133.00}{440}
\emline{138.00}{133.00}{441}{136.56}{131.88}{442}
\emline{136.56}{131.88}{443}{134.75}{131.00}{444}
\emline{134.75}{131.00}{445}{132.56}{130.38}{446}
\emline{132.56}{130.38}{447}{130.00}{130.00}{448}
\emline{130.00}{130.00}{449}{127.65}{130.07}{450}
\emline{127.65}{130.07}{451}{125.62}{129.83}{452}
\emline{125.62}{129.83}{453}{123.89}{129.29}{454}
\emline{123.89}{129.29}{455}{122.48}{128.43}{456}
\emline{122.48}{128.43}{457}{121.37}{127.26}{458}
\emline{121.37}{127.26}{459}{120.58}{125.79}{460}
\emline{120.58}{125.79}{461}{120.09}{124.00}{462}
\emline{120.09}{124.00}{463}{120.00}{120.00}{464}
\emline{120.00}{120.00}{465}{120.06}{117.88}{466}
\emline{120.06}{117.88}{467}{119.75}{116.00}{468}
\emline{119.75}{116.00}{469}{119.06}{114.38}{470}
\emline{119.06}{114.38}{471}{118.00}{113.00}{472}
\emline{118.00}{113.00}{473}{116.56}{111.88}{474}
\emline{116.56}{111.88}{475}{114.75}{111.00}{476}
\emline{114.75}{111.00}{477}{112.56}{110.38}{478}
\emline{112.56}{110.38}{479}{110.00}{110.00}{480}
\emline{110.00}{110.00}{481}{107.65}{110.07}{482}
\emline{107.65}{110.07}{483}{105.62}{109.83}{484}
\emline{105.62}{109.83}{485}{103.89}{109.29}{486}
\emline{103.89}{109.29}{487}{102.48}{108.43}{488}
\emline{102.48}{108.43}{489}{101.37}{107.26}{490}
\emline{101.37}{107.26}{491}{100.58}{105.79}{492}
\emline{100.58}{105.79}{493}{100.09}{104.00}{494}
\emline{100.09}{104.00}{495}{100.00}{100.00}{496}
\emline{100.00}{100.00}{497}{100.06}{97.88}{498}
\emline{100.06}{97.88}{499}{99.75}{96.00}{500}
\emline{99.75}{96.00}{501}{99.06}{94.38}{502}
\emline{99.06}{94.38}{503}{98.00}{93.00}{504}
\emline{98.00}{93.00}{505}{96.56}{91.88}{506}
\emline{96.56}{91.88}{507}{94.75}{91.00}{508}
\emline{94.75}{91.00}{509}{92.56}{90.38}{510}
\emline{92.56}{90.38}{511}{90.00}{90.00}{512}
\put(90.00,90.00){\circle*{3.00}}
\put(7.00,0.00){\makebox(0,0)[ct]{$W^+,\mu$}}
\put(180.00,0.00){\makebox(0,0)[ct]{$W^-,\lambda$}}
\put(7.00,180.00){\makebox(0,0)[cc]{$Z,\sigma$}}
\put(180.00,180.00){\makebox(0,0)[cc]{$Z,\nu$}}
\end{picture}
\emp
$- i g^2 \cos{\theta_W} V_{\nu\sigma\mu\lambda}$\hfill
\bmp{3.5cm}
\special{em:linewidth 0.3pt}
\unitlength 0.15mm
\linethickness{0.3pt}
\begin{picture}(171.00,159.00)
\emline{10.00}{10.00}{1}{12.35}{9.93}{2}
\emline{12.35}{9.93}{3}{14.38}{10.17}{4}
\emline{14.38}{10.17}{5}{16.11}{10.71}{6}
\emline{16.11}{10.71}{7}{17.52}{11.57}{8}
\emline{17.52}{11.57}{9}{18.63}{12.74}{10}
\emline{18.63}{12.74}{11}{19.42}{14.21}{12}
\emline{19.42}{14.21}{13}{19.91}{16.00}{14}
\emline{19.91}{16.00}{15}{20.00}{20.00}{16}
\emline{20.00}{20.00}{17}{19.94}{22.13}{18}
\emline{19.94}{22.13}{19}{20.25}{24.00}{20}
\emline{20.25}{24.00}{21}{20.94}{25.63}{22}
\emline{20.94}{25.63}{23}{22.00}{27.00}{24}
\emline{22.00}{27.00}{25}{23.44}{28.13}{26}
\emline{23.44}{28.13}{27}{25.25}{29.00}{28}
\emline{25.25}{29.00}{29}{27.44}{29.63}{30}
\emline{27.44}{29.63}{31}{30.00}{30.00}{32}
\emline{30.00}{30.00}{33}{32.35}{29.93}{34}
\emline{32.35}{29.93}{35}{34.38}{30.17}{36}
\emline{34.38}{30.17}{37}{36.11}{30.71}{38}
\emline{36.11}{30.71}{39}{37.52}{31.57}{40}
\emline{37.52}{31.57}{41}{38.63}{32.74}{42}
\emline{38.63}{32.74}{43}{39.42}{34.21}{44}
\emline{39.42}{34.21}{45}{39.91}{36.00}{46}
\emline{39.91}{36.00}{47}{40.00}{40.00}{48}
\emline{40.00}{40.00}{49}{39.94}{42.13}{50}
\emline{39.94}{42.13}{51}{40.25}{44.00}{52}
\emline{40.25}{44.00}{53}{40.94}{45.63}{54}
\emline{40.94}{45.63}{55}{42.00}{47.00}{56}
\emline{42.00}{47.00}{57}{43.44}{48.13}{58}
\emline{43.44}{48.13}{59}{45.25}{49.00}{60}
\emline{45.25}{49.00}{61}{47.44}{49.63}{62}
\emline{47.44}{49.63}{63}{50.00}{50.00}{64}
\emline{50.00}{50.00}{65}{52.35}{49.93}{66}
\emline{52.35}{49.93}{67}{54.38}{50.17}{68}
\emline{54.38}{50.17}{69}{56.11}{50.71}{70}
\emline{56.11}{50.71}{71}{57.52}{51.57}{72}
\emline{57.52}{51.57}{73}{58.63}{52.74}{74}
\emline{58.63}{52.74}{75}{59.42}{54.21}{76}
\emline{59.42}{54.21}{77}{59.91}{56.00}{78}
\emline{59.91}{56.00}{79}{60.00}{60.00}{80}
\emline{60.00}{60.00}{81}{59.94}{62.13}{82}
\emline{59.94}{62.13}{83}{60.25}{64.00}{84}
\emline{60.25}{64.00}{85}{60.94}{65.63}{86}
\emline{60.94}{65.63}{87}{62.00}{67.00}{88}
\emline{62.00}{67.00}{89}{63.44}{68.13}{90}
\emline{63.44}{68.13}{91}{65.25}{69.00}{92}
\emline{65.25}{69.00}{93}{67.44}{69.63}{94}
\emline{67.44}{69.63}{95}{70.00}{70.00}{96}
\emline{70.00}{70.00}{97}{72.35}{69.93}{98}
\emline{72.35}{69.93}{99}{74.38}{70.17}{100}
\emline{74.38}{70.17}{101}{76.11}{70.71}{102}
\emline{76.11}{70.71}{103}{77.52}{71.57}{104}
\emline{77.52}{71.57}{105}{78.63}{72.74}{106}
\emline{78.63}{72.74}{107}{79.42}{74.21}{108}
\emline{79.42}{74.21}{109}{79.91}{76.00}{110}
\emline{79.91}{76.00}{111}{80.00}{80.00}{112}
\emline{80.00}{80.00}{113}{79.94}{82.13}{114}
\emline{79.94}{82.13}{115}{80.25}{84.00}{116}
\emline{80.25}{84.00}{117}{80.94}{85.63}{118}
\emline{80.94}{85.63}{119}{82.00}{87.00}{120}
\emline{82.00}{87.00}{121}{83.44}{88.13}{122}
\emline{83.44}{88.13}{123}{85.25}{89.00}{124}
\emline{85.25}{89.00}{125}{87.44}{89.63}{126}
\emline{87.44}{89.63}{127}{90.00}{90.00}{128}
\emline{170.00}{10.00}{129}{170.07}{12.35}{130}
\emline{170.07}{12.35}{131}{169.83}{14.38}{132}
\emline{169.83}{14.38}{133}{169.29}{16.11}{134}
\emline{169.29}{16.11}{135}{168.43}{17.52}{136}
\emline{168.43}{17.52}{137}{167.26}{18.63}{138}
\emline{167.26}{18.63}{139}{165.79}{19.42}{140}
\emline{165.79}{19.42}{141}{164.00}{19.91}{142}
\emline{164.00}{19.91}{143}{160.00}{20.00}{144}
\emline{160.00}{20.00}{145}{157.88}{19.94}{146}
\emline{157.88}{19.94}{147}{156.00}{20.25}{148}
\emline{156.00}{20.25}{149}{154.38}{20.94}{150}
\emline{154.38}{20.94}{151}{153.00}{22.00}{152}
\emline{153.00}{22.00}{153}{151.88}{23.44}{154}
\emline{151.88}{23.44}{155}{151.00}{25.25}{156}
\emline{151.00}{25.25}{157}{150.38}{27.44}{158}
\emline{150.38}{27.44}{159}{150.00}{30.00}{160}
\emline{150.00}{30.00}{161}{150.07}{32.35}{162}
\emline{150.07}{32.35}{163}{149.83}{34.38}{164}
\emline{149.83}{34.38}{165}{149.29}{36.11}{166}
\emline{149.29}{36.11}{167}{148.43}{37.52}{168}
\emline{148.43}{37.52}{169}{147.26}{38.63}{170}
\emline{147.26}{38.63}{171}{145.79}{39.42}{172}
\emline{145.79}{39.42}{173}{144.00}{39.91}{174}
\emline{144.00}{39.91}{175}{140.00}{40.00}{176}
\emline{140.00}{40.00}{177}{137.88}{39.94}{178}
\emline{137.88}{39.94}{179}{136.00}{40.25}{180}
\emline{136.00}{40.25}{181}{134.38}{40.94}{182}
\emline{134.38}{40.94}{183}{133.00}{42.00}{184}
\emline{133.00}{42.00}{185}{131.88}{43.44}{186}
\emline{131.88}{43.44}{187}{131.00}{45.25}{188}
\emline{131.00}{45.25}{189}{130.38}{47.44}{190}
\emline{130.38}{47.44}{191}{130.00}{50.00}{192}
\emline{130.00}{50.00}{193}{130.07}{52.35}{194}
\emline{130.07}{52.35}{195}{129.83}{54.38}{196}
\emline{129.83}{54.38}{197}{129.29}{56.11}{198}
\emline{129.29}{56.11}{199}{128.43}{57.52}{200}
\emline{128.43}{57.52}{201}{127.26}{58.63}{202}
\emline{127.26}{58.63}{203}{125.79}{59.42}{204}
\emline{125.79}{59.42}{205}{124.00}{59.91}{206}
\emline{124.00}{59.91}{207}{120.00}{60.00}{208}
\emline{120.00}{60.00}{209}{117.88}{59.94}{210}
\emline{117.88}{59.94}{211}{116.00}{60.25}{212}
\emline{116.00}{60.25}{213}{114.38}{60.94}{214}
\emline{114.38}{60.94}{215}{113.00}{62.00}{216}
\emline{113.00}{62.00}{217}{111.88}{63.44}{218}
\emline{111.88}{63.44}{219}{111.00}{65.25}{220}
\emline{111.00}{65.25}{221}{110.38}{67.44}{222}
\emline{110.38}{67.44}{223}{110.00}{70.00}{224}
\emline{110.00}{70.00}{225}{110.07}{72.35}{226}
\emline{110.07}{72.35}{227}{109.83}{74.38}{228}
\emline{109.83}{74.38}{229}{109.29}{76.11}{230}
\emline{109.29}{76.11}{231}{108.43}{77.52}{232}
\emline{108.43}{77.52}{233}{107.26}{78.63}{234}
\emline{107.26}{78.63}{235}{105.79}{79.42}{236}
\emline{105.79}{79.42}{237}{104.00}{79.91}{238}
\emline{104.00}{79.91}{239}{100.00}{80.00}{240}
\emline{100.00}{80.00}{241}{97.88}{79.94}{242}
\emline{97.88}{79.94}{243}{96.00}{80.25}{244}
\emline{96.00}{80.25}{245}{94.38}{80.94}{246}
\emline{94.38}{80.94}{247}{93.00}{82.00}{248}
\emline{93.00}{82.00}{249}{91.88}{83.44}{250}
\emline{91.88}{83.44}{251}{91.00}{85.25}{252}
\emline{91.00}{85.25}{253}{90.38}{87.44}{254}
\emline{90.38}{87.44}{255}{90.00}{90.00}{256}
\emline{90.00}{90.00}{257}{90.00}{100.00}{258}
\emline{90.00}{105.00}{259}{90.00}{115.00}{260}
\emline{90.00}{120.00}{261}{90.00}{130.00}{262}
\emline{90.00}{135.00}{263}{90.00}{145.00}{264}
\emline{90.00}{150.00}{265}{90.00}{160.00}{266}
\put(90.00,90.00){\circle*{2.00}}
\put(6.00,0.00){\makebox(0,0)[ct]{$W^+,\mu$}}
\put(170.00,0.00){\makebox(0,0)[ct]{$W^-,\nu$}}
\put(115.00,160.00){\makebox(0,0)[cc]{$H$}}
\end{picture}
\emp
$ig m_W g_{\mu\nu}$\\[10mm]
\bmp{3.5cm}
\special{em:linewidth 0.3pt}
\unitlength 0.15mm
\linethickness{0.3pt}
\begin{picture}(171.00,159.00)
\emline{10.00}{10.00}{1}{12.35}{9.93}{2}
\emline{12.35}{9.93}{3}{14.38}{10.17}{4}
\emline{14.38}{10.17}{5}{16.11}{10.71}{6}
\emline{16.11}{10.71}{7}{17.52}{11.57}{8}
\emline{17.52}{11.57}{9}{18.63}{12.74}{10}
\emline{18.63}{12.74}{11}{19.42}{14.21}{12}
\emline{19.42}{14.21}{13}{19.91}{16.00}{14}
\emline{19.91}{16.00}{15}{20.00}{20.00}{16}
\emline{20.00}{20.00}{17}{19.94}{22.13}{18}
\emline{19.94}{22.13}{19}{20.25}{24.00}{20}
\emline{20.25}{24.00}{21}{20.94}{25.63}{22}
\emline{20.94}{25.63}{23}{22.00}{27.00}{24}
\emline{22.00}{27.00}{25}{23.44}{28.13}{26}
\emline{23.44}{28.13}{27}{25.25}{29.00}{28}
\emline{25.25}{29.00}{29}{27.44}{29.63}{30}
\emline{27.44}{29.63}{31}{30.00}{30.00}{32}
\emline{30.00}{30.00}{33}{32.35}{29.93}{34}
\emline{32.35}{29.93}{35}{34.38}{30.17}{36}
\emline{34.38}{30.17}{37}{36.11}{30.71}{38}
\emline{36.11}{30.71}{39}{37.52}{31.57}{40}
\emline{37.52}{31.57}{41}{38.63}{32.74}{42}
\emline{38.63}{32.74}{43}{39.42}{34.21}{44}
\emline{39.42}{34.21}{45}{39.91}{36.00}{46}
\emline{39.91}{36.00}{47}{40.00}{40.00}{48}
\emline{40.00}{40.00}{49}{39.94}{42.13}{50}
\emline{39.94}{42.13}{51}{40.25}{44.00}{52}
\emline{40.25}{44.00}{53}{40.94}{45.63}{54}
\emline{40.94}{45.63}{55}{42.00}{47.00}{56}
\emline{42.00}{47.00}{57}{43.44}{48.13}{58}
\emline{43.44}{48.13}{59}{45.25}{49.00}{60}
\emline{45.25}{49.00}{61}{47.44}{49.63}{62}
\emline{47.44}{49.63}{63}{50.00}{50.00}{64}
\emline{50.00}{50.00}{65}{52.35}{49.93}{66}
\emline{52.35}{49.93}{67}{54.38}{50.17}{68}
\emline{54.38}{50.17}{69}{56.11}{50.71}{70}
\emline{56.11}{50.71}{71}{57.52}{51.57}{72}
\emline{57.52}{51.57}{73}{58.63}{52.74}{74}
\emline{58.63}{52.74}{75}{59.42}{54.21}{76}
\emline{59.42}{54.21}{77}{59.91}{56.00}{78}
\emline{59.91}{56.00}{79}{60.00}{60.00}{80}
\emline{60.00}{60.00}{81}{59.94}{62.13}{82}
\emline{59.94}{62.13}{83}{60.25}{64.00}{84}
\emline{60.25}{64.00}{85}{60.94}{65.63}{86}
\emline{60.94}{65.63}{87}{62.00}{67.00}{88}
\emline{62.00}{67.00}{89}{63.44}{68.13}{90}
\emline{63.44}{68.13}{91}{65.25}{69.00}{92}
\emline{65.25}{69.00}{93}{67.44}{69.63}{94}
\emline{67.44}{69.63}{95}{70.00}{70.00}{96}
\emline{70.00}{70.00}{97}{72.35}{69.93}{98}
\emline{72.35}{69.93}{99}{74.38}{70.17}{100}
\emline{74.38}{70.17}{101}{76.11}{70.71}{102}
\emline{76.11}{70.71}{103}{77.52}{71.57}{104}
\emline{77.52}{71.57}{105}{78.63}{72.74}{106}
\emline{78.63}{72.74}{107}{79.42}{74.21}{108}
\emline{79.42}{74.21}{109}{79.91}{76.00}{110}
\emline{79.91}{76.00}{111}{80.00}{80.00}{112}
\emline{80.00}{80.00}{113}{79.94}{82.13}{114}
\emline{79.94}{82.13}{115}{80.25}{84.00}{116}
\emline{80.25}{84.00}{117}{80.94}{85.63}{118}
\emline{80.94}{85.63}{119}{82.00}{87.00}{120}
\emline{82.00}{87.00}{121}{83.44}{88.13}{122}
\emline{83.44}{88.13}{123}{85.25}{89.00}{124}
\emline{85.25}{89.00}{125}{87.44}{89.63}{126}
\emline{87.44}{89.63}{127}{90.00}{90.00}{128}
\emline{170.00}{10.00}{129}{170.07}{12.35}{130}
\emline{170.07}{12.35}{131}{169.83}{14.38}{132}
\emline{169.83}{14.38}{133}{169.29}{16.11}{134}
\emline{169.29}{16.11}{135}{168.43}{17.52}{136}
\emline{168.43}{17.52}{137}{167.26}{18.63}{138}
\emline{167.26}{18.63}{139}{165.79}{19.42}{140}
\emline{165.79}{19.42}{141}{164.00}{19.91}{142}
\emline{164.00}{19.91}{143}{160.00}{20.00}{144}
\emline{160.00}{20.00}{145}{157.88}{19.94}{146}
\emline{157.88}{19.94}{147}{156.00}{20.25}{148}
\emline{156.00}{20.25}{149}{154.38}{20.94}{150}
\emline{154.38}{20.94}{151}{153.00}{22.00}{152}
\emline{153.00}{22.00}{153}{151.88}{23.44}{154}
\emline{151.88}{23.44}{155}{151.00}{25.25}{156}
\emline{151.00}{25.25}{157}{150.38}{27.44}{158}
\emline{150.38}{27.44}{159}{150.00}{30.00}{160}
\emline{150.00}{30.00}{161}{150.07}{32.35}{162}
\emline{150.07}{32.35}{163}{149.83}{34.38}{164}
\emline{149.83}{34.38}{165}{149.29}{36.11}{166}
\emline{149.29}{36.11}{167}{148.43}{37.52}{168}
\emline{148.43}{37.52}{169}{147.26}{38.63}{170}
\emline{147.26}{38.63}{171}{145.79}{39.42}{172}
\emline{145.79}{39.42}{173}{144.00}{39.91}{174}
\emline{144.00}{39.91}{175}{140.00}{40.00}{176}
\emline{140.00}{40.00}{177}{137.88}{39.94}{178}
\emline{137.88}{39.94}{179}{136.00}{40.25}{180}
\emline{136.00}{40.25}{181}{134.38}{40.94}{182}
\emline{134.38}{40.94}{183}{133.00}{42.00}{184}
\emline{133.00}{42.00}{185}{131.88}{43.44}{186}
\emline{131.88}{43.44}{187}{131.00}{45.25}{188}
\emline{131.00}{45.25}{189}{130.38}{47.44}{190}
\emline{130.38}{47.44}{191}{130.00}{50.00}{192}
\emline{130.00}{50.00}{193}{130.07}{52.35}{194}
\emline{130.07}{52.35}{195}{129.83}{54.38}{196}
\emline{129.83}{54.38}{197}{129.29}{56.11}{198}
\emline{129.29}{56.11}{199}{128.43}{57.52}{200}
\emline{128.43}{57.52}{201}{127.26}{58.63}{202}
\emline{127.26}{58.63}{203}{125.79}{59.42}{204}
\emline{125.79}{59.42}{205}{124.00}{59.91}{206}
\emline{124.00}{59.91}{207}{120.00}{60.00}{208}
\emline{120.00}{60.00}{209}{117.88}{59.94}{210}
\emline{117.88}{59.94}{211}{116.00}{60.25}{212}
\emline{116.00}{60.25}{213}{114.38}{60.94}{214}
\emline{114.38}{60.94}{215}{113.00}{62.00}{216}
\emline{113.00}{62.00}{217}{111.88}{63.44}{218}
\emline{111.88}{63.44}{219}{111.00}{65.25}{220}
\emline{111.00}{65.25}{221}{110.38}{67.44}{222}
\emline{110.38}{67.44}{223}{110.00}{70.00}{224}
\emline{110.00}{70.00}{225}{110.07}{72.35}{226}
\emline{110.07}{72.35}{227}{109.83}{74.38}{228}
\emline{109.83}{74.38}{229}{109.29}{76.11}{230}
\emline{109.29}{76.11}{231}{108.43}{77.52}{232}
\emline{108.43}{77.52}{233}{107.26}{78.63}{234}
\emline{107.26}{78.63}{235}{105.79}{79.42}{236}
\emline{105.79}{79.42}{237}{104.00}{79.91}{238}
\emline{104.00}{79.91}{239}{100.00}{80.00}{240}
\emline{100.00}{80.00}{241}{97.88}{79.94}{242}
\emline{97.88}{79.94}{243}{96.00}{80.25}{244}
\emline{96.00}{80.25}{245}{94.38}{80.94}{246}
\emline{94.38}{80.94}{247}{93.00}{82.00}{248}
\emline{93.00}{82.00}{249}{91.88}{83.44}{250}
\emline{91.88}{83.44}{251}{91.00}{85.25}{252}
\emline{91.00}{85.25}{253}{90.38}{87.44}{254}
\emline{90.38}{87.44}{255}{90.00}{90.00}{256}
\emline{90.00}{90.00}{257}{90.00}{100.00}{258}
\emline{90.00}{105.00}{259}{90.00}{115.00}{260}
\emline{90.00}{120.00}{261}{90.00}{130.00}{262}
\emline{90.00}{135.00}{263}{90.00}{145.00}{264}
\emline{90.00}{150.00}{265}{90.00}{160.00}{266}
\put(90.00,90.00){\circle*{2.00}}
\put(6.00,0.00){\makebox(0,0)[ct]{$Z,\mu$}}
\put(170.00,0.00){\makebox(0,0)[ct]{$Z,\nu$}}
\put(115.00,160.00){\makebox(0,0)[cc]{$H$}}
\end{picture}
\emp
$ig\frac{m_Z}{\cos{\theta_W}} g_{\mu\nu}$\\[1cm]
where
\be V_{\lambda\mu\nu}(k,p,q) = (k-p)_{\nu} \,g_{\lambda\mu}
+ (p-q)_{\lambda}\, g_{\mu\nu} + (q-k)_{\mu}\,
g_{\lambda\nu},\quad k+p+q=0
\label{3v}
\ee
\[V_{\mu\nu\rho\sigma} = 2 g_{\mu\nu} g_{\rho\sigma} -
g_{\mu\rho} g_{\nu\sigma} - g_{\nu\rho} g_{\mu\sigma} \ .\]
Propagators of the gauge  bosons are
\[D^{\alpha\beta}_V(q) = \frac{-g^{\alpha\beta} + \frac{q^{\alpha}
q^{\beta}}{m_V^2}}{q^2 -m_V^2} \equiv \frac{P^{\alpha\beta}_V(q)}{q^2
-m_V^2}, \quad D^{\alpha\beta}_{\gamma} = \frac{-g^{\alpha\beta}}{q^2}
\equiv \frac{P^{\alpha\beta}_{\gamma}}{q^2} \ .\]
Mandelstam variables are defined as
\ber
s &=& (k_1 + k_2)^2 = (k_3 + k_4)^2 \nn\\
t &=& (k_1 -k_3)^2 = (k_4 -k_2)^2 \nn\\
u &=& (k_1 -k_4)^2 = (k_3 -k_2)^2 \nn
\eer
The expression containing polarization vectors has in all
amplitudes the form
\[
{\cal E}^{\mu\nu\rho\sigma} =
\varepsilon^{\mu}_1\,\varepsilon^{\nu}_2\,
 \varepsilon^{*\rho}_3\,\varepsilon^{*\sigma}_4
\]
where $\ep_i = \ep(k_i)$. I use the abbreviation $x =
\cos{\theta_{cm}}$, where $\theta_{cm}$ is the angle between $\bk_1$ a
$\bk_3$ in  CMS.

\subsection{$W^+(k_1) + W^-(k_2) \to Z(k_3) + Z(k_4)$}
\[\M^{(1)} = - g^2\cos^2{\theta_W}\left[\frac{A^{(1)}_{tW}}{t -
m_W^2} + \frac{A^{(1)}_{uW}}{u - m_W^2}  + A^{(1)}_c\right] +
\M^{(1)}_{sH}
\]
\[A^{(1)}_{tW}=
 V_{\mu\alpha\rho}(-k_1,q,k_3)\,V_{\beta\nu\sigma}(-q,-k_2,k_4)
 P^{\alpha\beta}_W(q){\cal E}^{\mu\nu\rho\sigma}
\]
\[A^{(1)}_{uW}=
 V_{\mu\alpha\sigma}(-k_1,q,k_4)\,V_{\beta\nu\rho}(-q,-k_2,k_3)
P^{\alpha\beta}_W(q){\cal E}^{\mu\nu\rho\sigma}
\]
\[A^{(1)}_c = V_{\rho\sigma\mu\nu}{\cal E}^{\mu\nu\rho\sigma}\]
After contraction
\ber
A^{(1)}_{tW} &=&
      4\,(k_1\cdot\varepsilon_3^*)\,(k_2\cdot\varepsilon_4^*)\,
      (\varepsilon_1\cdot\varepsilon_2) +
      4\,(k_1\cdot\varepsilon_3^*)\,(k_4\cdot\varepsilon_2)\,
      (\varepsilon_1\cdot\varepsilon_4^*)\nn\\
&+&   4\,(k_2\cdot\varepsilon_4^*)\,(k_3\cdot\varepsilon_1)\,
      (\varepsilon_2\cdot\varepsilon_3^*)+
      4\,(k_3\cdot\varepsilon_1)\,(k_4\cdot\varepsilon_2)\,
      (\varepsilon_3^*\cdot\varepsilon_4^*) \nn\\
&-&  2\,(k_4\cdot\varepsilon_2)\,
     ((k_1 + k_3)\cdot \varepsilon_4^*)\,(\varepsilon_1\cdot\varepsilon_3^*)
     - 2\,(k_2\cdot\varepsilon_4^*)\,
     \left((k_1 + k_3)\cdot\varepsilon_2\right) \,
     (\varepsilon_1\cdot\varepsilon_3^*) \nn\\
&-&   2\,(k_1\cdot\varepsilon_3^*)\,((k_2 +k_4)\cdot\varepsilon_1)\,
     (\varepsilon_2\cdot\varepsilon_4^*)- 2\,(k_3\cdot\varepsilon_1)\,
     \left((k_2 +k_4)\cdot\varepsilon_3^*\right) \,
     (\varepsilon_2\cdot\varepsilon_4^*)\nn\\
&+&  \frac{(m_W^2 - m_Z^2)^2}{m_W^2}\,(\varepsilon_1\cdot\varepsilon_3^*)\,
     (\varepsilon_2\cdot\varepsilon_4^*)+
     (s - u) \,(\varepsilon_1\cdot\varepsilon^*_3)\,
     (\varepsilon_2\cdot\varepsilon_4^*).
\eer
\[ A^{(1)}_{uW} = A^{(1)}_{tW} (3\leftrightarrow 4, u\to t) \]
\[ A^{(1)}_c = 2\,(\varepsilon_1\cdot
\varepsilon_2)\,(\varepsilon^*_3\cdot \varepsilon^*_4) -
(\varepsilon_1\cdot
\varepsilon^*_4)\,(\varepsilon_2\cdot\varepsilon^*_3) -
(\varepsilon_1\cdot \varepsilon^*_3)\,(\varepsilon_2\cdot
\varepsilon^*_4)
\]
Replacing all polarization vectors $\varepsilon_i$ by $\varepsilon_i(L)$
given in (\ref{longpolvec}) we get amplitudes for longitudinally
polarized gauge bosons in CMS 
\ber A^{(1)}_{tW}(s,x)&=&
\frac{1}{32\,m_W^4\,m_Z^2}\,[-96\,m_W^4\,m_Z^4 +
    32\,m_W^2\,m_Z^6 + 8\,m_W^4\,m_Z^2\,s + 16\,m_W^2\,m_Z^4\,s\nn\\
&-& 8\,m_Z^6\,s - 4\,m_W^4\,s^2 - 10\,m_W^2\,m_Z^2\,s^2 + 2\,m_Z^4\,s^2 +
    3\,m_W^2\,s^3\nn\\
&+& 16\,m_W^4\,m_Z^2\,\beta_W\,\beta_Z\,s\,x
    + 12\,m_W^4\,\beta_W\,\beta_Z\,s^2\,x
    +24\,m_W^2\,m_Z^2\,\beta_W\,\beta_Z\,s^2\,x\nn\\
&-& 4\,m_Z^4\,\beta_W\,\beta_Z\,s^2\,x - 5\,m_W^2\,\beta_W\,\beta_Z\,s^3\,x
    + 32\,m_W^6\,s\,x^2\nn\\
&+& 96\,m_W^4\,m_Z^2\,s\,x^2 + 32\,m_W^2\,m_Z^4\,s\,x^2 -
    16\,m_W^4\,s^2\,x^2\nn\\
&-& 22\,m_W^2\,m_Z^2\,s^2\,x^2 + 2\,m_Z^4\,s^2\,x^2 +
    m_W^2\,s^3\,x^2 + m_W^2\,\beta_W\,
    \beta_Z\,s^3\,x^3\,]
\eer
\[A^{(1)}_{uW}(s,x) = A^{(1)}_{tW}(s,-x)\]
and
\[
A^{(1)}_c =\frac{s\,(-4\,m_W^2 - 4\,m_Z^2 + 3\,s - s\,x^2)}{8\,m_W^2\,m_Z^2}
\]
Kinematical variables are related by
\ber t &=& m_W^2 + m_Z^2 - \frac{s}{2} + \frac{s}{2}\,\beta_W\beta_Z \cos{\theta_{cm}}\nn\\
     u &=& m_W^2 + m_Z^2 - \frac{s}{2} - \frac{s}{2}\,\beta_W\beta_Z
     \cos{\theta_{cm}}\nn
\eer
where    
\[\beta_W = \sqrt{1 - \frac{4 m_W^2}{s}}\quad \beta_Z = \sqrt{1 - \frac{4 m_Z^2}{s}}
\]
\[ \M^{(1)}_{gauge} =
\frac{g^2\,m_Z^2\,\cos{\theta_W}^2}{4\,m_W^4}\,\frac{C^{(1)}}{J^{(1)}} \]
\ber  C^{(1)} &=&
       96\,{m_W^4}\,{m_Z^2} -
       32\,{m_W^2}\,{m_Z^4} -
       48\,{m_W^4}\,s +
       8\,{m_W^2}\,{m_Z^2}\,s\nn\\ &+&
       8\,{m_Z^4}\,s + 4\,{m_W^2}\,{s^2} -
       6\,{m_Z^2}\,{s^2} + {s^3} +
       128\,{m_W^6}\,{x^2} +
       32\,{m_W^4}\,s\,{x^2}\nn\\ &-&
       64\,{m_W^2}\,{m_Z^2}\,s\,{x^2} +
       8\,{m_W^2}\,{s^2}\,{x^2} +
       6\,{m_Z^2}\,{s^2}\,{x^2} - {s^3}\,{x^2}\nn
\eer
\ber    J^{(1)} &=&
       4\,{m_Z^4} - 4\,{m_Z^2}\,s + {s^2} -
       16\,{m_W^2}\,{m_Z^2}\,{x^2}\nn\\ &+&
       4\,{m_W^2}\,s\,{x^2} +
       4\,{m_Z^2}\,s\,{x^2} - {s^2}\,{x^2}\nn
\eer
For the Higgs boson amplitude (\ref{m1sh}) we have
\[\M^{(1)}_{sH} = - \frac{g^2 m_W m_Z}{\cos{\theta_W}}
\frac{(2 m_W^2 - s)\,(2 m_Z^2 - s)}{4 m_W^2 m_Z^2 (s - m_H^2 + i
m_H\Gamma_H)} \]

\subsection{$W^+(k_1) + Z(k_2) \to Z(k_3) + W^+(k_4)$}
\[\M^{(2)} = - g^2\cos^2{\theta_W}\left[\frac{A^{(2)}_{sW}}{s -
m_W^2} + \frac{A^{(2)}_{tW}}{t - m_W^2}  + A^{(2)}_c\right] +
\M^{(2)}_{uH}
\]
\[A^{(2)}_{sW}=
 V_{\mu\alpha\nu}(-k_1,q,-k_2)\,V_{\beta\sigma\rho}(-q,k_4,k_3)
 P^{\alpha\beta}_W(q){\cal E}^{\mu\nu\rho\sigma}
\]
\[A^{(2)}_{tW}=
 V_{\mu\alpha\rho}(-k_1,q,k_3)\,V_{\beta\sigma\nu}(-q,k_4,-k_2)
 P^{\alpha\beta}_W(q){\cal E}^{\mu\nu\rho\sigma}
\]
\[A^{(2)}_c = V_{\mu\sigma\nu\rho}{\cal E}^{\mu\nu\rho\sigma}\]
After contraction
\ber
A^{(2)}_{sW} &=& 4\,(k_1\cdot \varepsilon_2)\,(k_3 \cdot
\varepsilon^*_4)\,(\varepsilon_1 \cdot \varepsilon^*_3) -
4\,(k1\cdot\varepsilon_2)\,(k_4\cdot\varepsilon^*_3)\,(\varepsilon_1
\cdot\varepsilon^*_4)\nn\\
&-& 4\,(k_2\cdot\varepsilon_1)\,(k_3\cdot\varepsilon^*_4)\,
(\varepsilon_2\cdot\varepsilon^*_3) + 4\,(k_2\cdot\varepsilon_1)\,
(k_4\cdot\varepsilon^*_3)\,(\varepsilon_2 \cdot\varepsilon^*_4)\nn\\
&-& 2 \, (k_3 \cdot \varepsilon^*_4)\,(\varepsilon_1 \cdot
\varepsilon_2)\,((k_1 - k_2)\cdot \varepsilon^*_3) + 2\,(k_4 \cdot
\varepsilon^*_3)\,(\varepsilon_1 \cdot \varepsilon_2)\,((k_1 - k_2)
\cdot \varepsilon^*_4)\nn\\
&-& 2 \,(k_1 \cdot \varepsilon_2)\,(\varepsilon^*_3
\cdot \varepsilon^*_4)\,((k_3 - k_4) \cdot \varepsilon_1) + 2 \,(k_2
\cdot \varepsilon_1)\,(\varepsilon^*_3 \cdot \varepsilon^*_4)\,((k_3
- k_4) \cdot \varepsilon_2)\nn\\
&+& \frac{(m_W^2 - m_Z^2)^2}{m_W^2}\,(\varepsilon_1\cdot
\varepsilon_2)\,(\varepsilon^*_3\cdot\varepsilon^*_4) + (u - t)
\,(\varepsilon_1 \cdot \varepsilon_2)\,(\varepsilon^*_3 \cdot
\varepsilon^*_4)
\eer
\[A^{(2)}_{tW} = - A^{(2)}_{sW}(k_2 \leftrightarrow -k_3,
\ep_2 \leftrightarrow \ep_3^*) \]

\[
 A^{(2)}_c = 2\,(\varepsilon_1\cdot
 \varepsilon^*_4)\,(\varepsilon_2\cdot \varepsilon^*_3) -
 (\varepsilon_1\cdot \varepsilon^*_3)\,(\varepsilon_2\cdot
 \varepsilon^*_4) - (\varepsilon_1\cdot
 \varepsilon_2)\,(\varepsilon^*_3\cdot \varepsilon^*_4)
\]
\[t = m_W^2 + m_Z^2 -\frac{s}{2} + \frac{(m_Z^2 - m_W^2)^2}{2 s} + 2\,k^2
\cos{\theta_{cm}} \]
\[u = - 2 k^2 (1 + \cos{\theta_{cm}}) \]
where in CMS
\[ k^2 = \frac{1}{4 s}\left [s^2 + (m_W^2 - m_Z^2)^2 - 2s
\,(m_W^2 +m_Z^2)\right ] = |\bk_i|^2 \quad i = 1,2,3,4 \]
\[ \M^{(2)}_{gauge} =
\frac{g^2\,m_Z^2\,\cos{\theta_W}^2}{8\,m_W^4\,s\,
     \left(s - {m_W^2}\right)}\,\frac{C^{(2)}}{J^{(2)}}
\]

\ber C^{(2)} &=& 3\,{m_W^{10}} -
       12\,{m_W^8}\,{m_Z^2} +
       18\,{m_W^6}\,{m_Z^4} -
       12\,{m_W^4}\,{m_Z^6} +
       3\,{m_W^2}\,{m_Z^8} \nn\\ &+&
       17\,{m_W^8}\,s -
       32\,{m_W^6}\,{m_Z^2}\,s +
       10\,{m_W^4}\,{m_Z^4}\,s +
       8\,{m_W^2}\,{m_Z^6}\,s \nn\\ &-&
       3\,{m_Z^8}\,s + 26\,{m_W^6}\,{s^2} +
       32\,{m_W^4}\,{m_Z^2}\,{s^2} -
       30\,{m_W^2}\,{m_Z^4}\,{s^2} \nn\\ &+&
       4\,{m_Z^6}\,{s^2} - 50\,{m_W^4}\,{s^3} +
       16\,{m_W^2}\,{m_Z^2}\,{s^3} +
       2\,{m_Z^4}\,{s^3} + 3\,{m_W^2}\,{s^4} \nn\\ &-&
       4\,{m_Z^2}\,{s^4} + {s^5} +
       6\,{m_W^{10}}\,x -
       24\,{m_W^8}\,{m_Z^2}\,x +
       36\,{m_W^6}\,{m_Z^4}\,x \nn\\ &-&
       24\,{m_W^4}\,{m_Z^6}\,x +
       6\,{m_W^2}\,{m_Z^8}\,x -
       16\,{m_W^8}\,s\,x +
       36\,{m_W^6}\,{m_Z^2}\,s\,x\nn\\ &-&
       28\,{m_W^4}\,{m_Z^4}\,s\,x +
       12\,{m_W^2}\,{m_Z^6}\,s\,x -
       4\,{m_Z^8}\,s\,x + 20\,{m_W^6}\,{s^2}\,x\nn\\ &-&
       44\,{m_W^4}\,{m_Z^2}\,{s^2}\,x -
       20\,{m_W^2}\,{m_Z^4}\,{s^2}\,x +
       12\,{m_Z^6}\,{s^2}\,x -
       16\,{m_W^4}\,{s^3}\,x \nn\\ &-&
       4\,{m_W^2}\,{m_Z^2}\,{s^3}\,x -
       12\,{m_Z^4}\,{s^3}\,x +
       6\,{m_W^2}\,{s^4}\,x +
       4\,{m_Z^2}\,{s^4}\,x +
       3\,{m_W^{10}}\,{x^2}\nn\\ &-&
       12\,{m_W^8}\,{m_Z^2}\,{x^2} +
       18\,{m_W^6}\,{m_Z^4}\,{x^2} -
       12\,{m_W^4}\,{m_Z^6}\,{x^2} +
       3\,{m_W^2}\,{m_Z^8}\,{x^2}\nn\\ &-&
       33\,{m_W^8}\,s\,{x^2} +
       68\,{m_W^6}\,{m_Z^2}\,s\,{x^2} -
       38\,{m_W^4}\,{m_Z^4}\,s\,{x^2} +
       4\,{m_W^2}\,{m_Z^6}\,s\,{x^2}\nn\\ &-&
       {m_Z^8}\,s\,{x^2} +
       122\,{m_W^6}\,{s^2}\,{x^2} +
       12\,{m_W^4}\,{m_Z^2}\,{s^2}\,{x^2} -
       6\,{m_W^2}\,{m_Z^4}\,{s^2}\,{x^2}\nn\\ &+&
       34\,{m_W^4}\,{s^3}\,{x^2} -
       4\,{m_W^2}\,{m_Z^2}\,{s^3}\,{x^2} +
       2\,{m_Z^4}\,{s^3}\,{x^2} +
       3\,{m_W^2}\,{s^4}\,{x^2} - {s^5}\,{x^2}\nn
\eer
\ber  J^{(2)} &=& {m_W^4} -
       2\,{m_W^2}\,{m_Z^2} + {m_Z^4} +
       2\,{m_Z^2}\,s - {s^2} + {m_W^4}\,x -
       2\,{m_W^2}\,{m_Z^2}\,x\nn\\ &+&
       {m_Z^4}\,x - 2\,{m_W^2}\,s\,x -
       2\,{m_Z^2}\,s\,x + {s^2}\,x\nn
\eer
Higgs boson amplitude (\ref{m2uh})

\[\M^{(2)}_{uH} = \frac{g^2}{8\,m_W\,m_Z\,{\cos{\theta_W}}s}\,\frac{C^{(2)}_{uH}}{J^{(2)}_{uH}}\]
\ber
C^{(2)}_{uH} &=&{m_W^8} - 4\,{m_W^6}\,{m_Z^2} +
       6\,{m_W^4}\,{m_Z^4} -
       4\,{m_W^2}\,{m_Z^6} + {m_Z^8} -
       4\,{m_W^6}\,s + 4\,{m_W^4}\,{m_Z^2}\,s\nn\\ &+&
       4\,{m_W^2}\,{m_Z^4}\,s - 4\,{m_Z^6}\,s +
       6\,{m_W^4}\,{s^2} + 4\,{m_W^2}\,{m_Z^2}\,{s^2} +
       6\,{m_Z^4}\,{s^2} - 4\,{m_W^2}\,{s^3}\nn\\ &-&
       4\,{m_Z^2}\,{s^3} + {s^4} + 2\,{m_W^8}\,x -
       8\,{m_W^6}\,{m_Z^2}\,x+
       12\,{m_W^4}\,{m_Z^4}\,x\nn\\&-&
       8\,{m_W^2}\,{m_Z^6}\,x+
       2\,{m_Z^8}\,x -
       4\,{m_W^6}\,s\,x + 4\,{m_W^4}\,{m_Z^2}\,s\,x+
       4\,{m_W^2}\,{m_Z^4}\,s\,x - 4\,{m_Z^6}\,s\,x\nn\\ &+&
       4\,{m_W^4}\,{s^2}\,x-
       8\,{m_W^2}\,{m_Z^2}\,{s^2}\,x+
       4\,{m_Z^4}\,{s^2}\,x\nn\\ &-&
       4\,{m_W^2}\,{s^3}\,x -
       4\,{m_Z^2}\,{s^3}\,x + 2\,{s^4}\,x + {m_W^8}\,{x^2} -
       4\,{m_W^6}\,{m_Z^2}\,{x^2}+
       6\,{m_W^4}\,{m_Z^4}\,{x^2}\nn\\&-&
       4\,{m_W^2}\,{m_Z^6}\,{x^2}+
       {m_Z^8}\,{x^2} -
       2\,{m_W^4}\,{s^2}\,{x^2} +
       4\,{m_W^2}\,{m_Z^2}\,{s^2}\,{x^2} -
       2\,{m_Z^4}\,{s^2}\,{x^2} + {s^4}\,{x^2}\nn
\eer

\ber
J^{(2)}_{uH}&=& {m_W^4} -
       2\,{m_W^2}\,{m_Z^2} + {m_Z^4} +
       2\,{m_H^2}\,s - 2\,{m_W^2}\,s - 2\,{m_Z^2}\,s +
       {s^2} + {m_W^4}\,x\nn\\&-&
       2\,{m_W^2}\,{m_Z^2}\,x+
       {m_Z^4}\,x - 2\,{m_W^2}\,s\,x -
       2\,{m_Z^2}\,s\,x + {s^2}\,x\nn
\eer

\subsection{$W^+(k_1) + W^+(k_2) \to W^+(k_3) + W^+(k_4)$}
\[\M^{(3)} = - g^2\cos^2{\theta_W}\left[\frac{A^{(3)}_{tZ}}{t -
m_Z^2} + \frac{A^{(3)}_{uZ}}{u - m_Z^2}\right] - g^2\sin^2{\theta_W}
\left[\frac{A^{(3)}_{t\gamma}}{t} + \frac{A^{(3)}_{u\gamma}}{u}\right] +
g^2 A^{(3)}_c + \M^{(3)}_{tH} + \M^{(3)}_{uH}\ . \]

\[A^{(3)}_{tZ,\gamma}=
 V_{\mu\rho\alpha}(-k_1,k_3,q)\,V_{\nu\sigma\beta}(-k_2,k_4,-q,)
      P^{\alpha\beta}_{Z,\gamma}(q){\cal E}^{\mu\nu\rho\sigma}
\]
\[A^{(3)}_{uZ,\gamma}=
V_{\mu\sigma\alpha}(-k_1,k_4,q)\,V_{\nu\rho\beta}(-k_2,k_3,-q)
  P^{\alpha\beta}_{Z,\gamma}(q){\cal E}^{\mu\nu\rho\sigma}
\]

\[A^{(3)}_c = V_{\mu\nu\rho\sigma}{\cal E}^{\mu\nu\rho\sigma}\ .\]
Kinematical variables are given by
\[t = 2 \left( m_W^2 - \frac{s}{4}\right) (1 - x) \qquad
  u = 2 \left( m_W^2 - \frac{s}{4}\right) (1 + x) \]
\ber
A^{(3)}_{tZ} &=&2\,(k_2\cdot\varepsilon^*_4)\,
        \left((k_1\cdot\varepsilon_2) + (k_3\cdot\varepsilon_2) \right) \,
        (\varepsilon_1\cdot\varepsilon^*_3) - 4\,(k_1\cdot\varepsilon^*_3)\,(k_2\cdot \varepsilon^*_4)\,
        (\varepsilon_1\cdot\varepsilon_2) \nn\\
 &+&   2\,\left(  (k_1\cdot\varepsilon^*_4) + (k_3\cdot\varepsilon^*_4) \right) \,
       (k_4\cdot\varepsilon_2)\,(\varepsilon_1\cdot\varepsilon^*_3) -
       4\, (k_1\cdot\varepsilon^*_3)\,(k_4\cdot\varepsilon_2)\,
       (\varepsilon_1\cdot\varepsilon^*_4)\nn\\
 &+&  2\, (k_1\cdot\varepsilon^*_3)\,\left((k_2\cdot \varepsilon_1)
    + (k_4\cdot\varepsilon_1) \right)\,(\varepsilon_2\cdot\varepsilon^*_4)
    - 4\, (k_2\cdot\varepsilon^*_4)\,(k_3\cdot\varepsilon_1)\,
        (\varepsilon_2\cdot\varepsilon^*_3)\nn\\
&+&  2\, (k_3\cdot\varepsilon_1)\,\left((k_2\cdot\varepsilon^*_3) +
      (k_4\cdot\varepsilon^*_3)
     \right)\,(\varepsilon_2\cdot\varepsilon^*_4)\nn\\
&-&  \left( s - u \right)\,(\varepsilon_1\cdot\varepsilon^*_3)\,
    (\varepsilon_2\cdot\varepsilon^*_4) - 4\,(k_3\cdot\varepsilon_1)\,
    (k_4\cdot\varepsilon_2)\,(\varepsilon^*_3\cdot\varepsilon^*_4)\ .
\eer
\[A^{(3)}_{uZ} = A^{(3)}_{tZ} (3\leftrightarrow 4)\]
\[A^{(3)}_{(t,u)\gamma} = A^{(3)}_{(t,u)Z}\]
For relations among different $A$s see table\,2 and below.
\ber A^{(3)}_{tZ}(s,x)=\frac{1}{(32\,m_W^4)}\,&(&64\,m_W^6 - 16\,m_W^4\,s + 12\,m_W^2\,s^2 - 3\,s^3 + 64\,m_W^6\,x\nn\\
&+& 112\,m_W^4\,s\,x - 52\,m_W^2\,s^2\,x + 5\,s^3\,x -
160\,m_W^4\,s\,x^2\nn\\
&+& 36\,m_W^2\,s^2\,x^2 - s^3\,x^2 + 4\,m_W^2\,s^2\,x^3 - s^3\,x^3)
\label{a3tz}
\eer
\[ A^{(3)}_{uZ}(s,x) = A^{(3)}_{tZ}(s,-x)\]
\[
\M^{(3)}_{gauge} = {\frac{{g^2}\,s\,\left( -8\,{m_W^2} + 3\,s - s\,{x^2} \right) }
    {8\,{m_W^4}}} +
    \frac{g^2\,\cos^2{\theta_W}}{8\,{m_W^4}}\,\frac{C^{(3)}_Z}{J^{(3)}_Z}
    -
    \frac{g^2\,\sin^2{\theta_W}}{8\,{m_W^4}}\,
    \frac{C^{(3)}_{\gamma}}{J^{(3)}_{\gamma}}
\]
\ber C^{(3)}_Z &=& 256\,{m_W^8} -
        128\,{m_W^6}\,{m_Z^2} -
        128\,{m_W^6}\,s +
        32\,{m_W^4}\,{m_Z^2}\,s +
        64\,{m_W^4}\,{s^2}\nn\\ &-&
        24\,{m_W^2}\,{m_Z^2}\,{s^2} -
        24\,{m_W^2}\,{s^3} + 6\,{m_Z^2}\,{s^3} +
        3\,{s^4} + 256\,{m_W^8}\,{x^2}\nn\\ &-&
        256\,{m_W^6}\,s\,{x^2} +
        320\,{m_W^4}\,{m_Z^2}\,s\,{x^2} -
        16\,{m_W^4}\,{s^2}\,{x^2}\nn\\ &-&
        72\,{m_W^2}\,{m_Z^2}\,{s^2}\,{x^2} +
        32\,{m_W^2}\,{s^3}\,{x^2} +
        2\,{m_Z^2}\,{s^3}\,{x^2}\nn\\ &-&
        4\,{s^4}\,{x^2} +
        16\,{m_W^4}\,{s^2}\,{x^4} -
        8\,{m_W^2}\,{s^3}\,{x^4} + {s^4}\,{x^4}\nn
\eer
\[
J^{(3)}_Z = \left( 4\,{m_W^2} - 2\,{m_Z^2} - s -
        4\,{m_W^2}\,x + s\,x \right) \,
      \left( -4\,{m_W^2} + 2\,{m_Z^2} + s -
        4\,{m_W^2}\,x + s\,x \right)
\]
\ber
C^{(3)}_{\gamma} &=& 64\,{m_W^6} -
        16\,{m_W^4}\,s + 12\,{m_W^2}\,{s^2} -
        3\,{s^3} + 64\,{m_W^6}\,{x^2}\nn\\ &-&
        48\,{m_W^4}\,s\,{x^2} -
        16\,{m_W^2}\,{s^2}\,{x^2} + 4\,{s^3}\,{x^2} +
        4\,{m_W^2}\,{s^2}\,{x^4} - {s^3}\,{x^4}\nn
\eer
\[
J^{(3)}_{\gamma} =\left( s - 4\,{m_W^2} \right) \,\left( {x^2} - 1 \right)
\]
Higgs boson amplitude (\ref{m3h})
\[\M^{(3)}_H = g^2\,\frac{C^{(3)}_H}{J^{(3)}_H}\]
\ber C^{(3)}_H &=& 32\,{m_H^2}\,{m_W^4} - 64\,{m_W^6} -
       16\,{m_H^2}\,{m_W^2}\,s + 48\,{m_W^4}\,s\nn\\ &+&
       2\,{m_H^2}\,{s^2} - 12\,{m_W^2}\,{s^2} + {s^3} -
       32\,{m_W^4}\,s\,{x^2} + 2\,{m_H^2}\,{s^2}\,{x^2}\nn\\ &+&
       12\,{m_W^2}\,{s^2}\,{x^2} - {s^3}\,{x^2}
\eer
\[J^{(3)}_H =  4\,{m_W^2}\,\left( 2\,{m_H^2} - 4\,{m_W^2} + s +
       4\,{m_W^2}\,x - s\,x \right) \,
     \left( 2\,{m_H^2} - 4\,{m_W^2} + s -
       4\,{m_W^2}\,x + s\,x \right)
\]

\subsection{$W^+(k_1) + W^-(k_2) \to W^+(k_3) + W^-(k_4)$}
\[\M^{(4)} = - g^2\cos^2{\theta_W}\left[\frac{A^{(4)}_{sZ}}{s -
m_Z^2} + \frac{A^{(4)}_{tZ}}{t - m_Z^2}\right] - g^2\sin^2{\theta_W}
\left[\frac{A^{(4)}_{s\gamma}}{s} + \frac{A^{(4)}_{t\gamma}}{t}\right] +
g^2 A^{(4)}_c + \M^{(4)}_{sH} + \M^{(4)}_{tH}\ . \]
\[A^{(4)}_{sZ,\gamma}=
  V_{\mu\nu\alpha}(-k_1,-k_2,q)\,V_{\sigma\rho\beta}(k_4,k_3,-q,)
  P^{\alpha\beta}_{Z,\gamma}(q){\cal E}^{\mu\nu\rho\sigma}
\]
\[A^{(4)}_{tZ,\gamma}=
 V_{\mu\rho\alpha}(-k_1,k_3,q)\,V_{\sigma\nu\beta}(k_4,-k_2,-q)
      P^{\alpha\beta}_{Z,\gamma}(q){\cal E}^{\mu\nu\rho\sigma}
\]
\[A^{(4)}_c = g^2
V_{\mu\sigma\rho\nu}{\cal E}^{\mu\nu\rho\sigma}\ .\]
\ber \M^{(4)}_{gauge} &=& {\frac{{g^2}\,s\,\left( 8\,{m_W^2} - 3\,s -
        24\,{m_W^2}\,x + 6\,s\,x + s\,{x^2} \right) }{16\,
      {m_W^4}}}\nn\\ &+&
      \M^{(4)}_{sZ} + \M^{(4)}_{s\gamma}- {g^2}\,{\cos^2{\theta_W}}\,\frac{C^{(4)}_{tZ}}
      {J^{(4)}_{tZ}}  - {g^2}\,{\sin^2{\theta_W}}\,
      \frac{C^{(4)}_{t\gamma}}{J^{(4)}_{t\gamma}}\nn
\eer
\[
\M^{(4)}_{sZ} =  - g^2\,\cos^2{\theta_W}\,\frac{\left( 4\,{m_W^2} - s \right) \,
         \left( 2\,{m_W^2} + s \right)^2\,x}{4\,
         {m_W^4}\,\left( {m_Z^2} - s \right)}
\]
\[
\M^{(4)}_{s\gamma} =  - {g^2}\,\sin^2{\theta_W}\left( -3\,x -
\frac{4\,{m_W^2}\,x}{s} + \frac{s^2\,x}{4\,m_W^4}\right)
\]
\ber
C^{(4)}_{tZ} &=& -64\,{m_W^6} + 16\,{m_W^4}\,s -
         12\,{m_W^2}\,{s^2} + 3\,{s^3} -
         64\,{m_W^6}\,x - 112\,{m_W^4}\,s\,x +
         52\,{m_W^2}\,{s^2}\,x\nn\\ &-&
         5\,{s^3}\,x +
         160\,{m_W^4}\,s\,{x^2} -
         36\,{m_W^2}\,{s^2}\,{x^2} + {s^3}\,{x^2} -
         4\,{m_W^2}\,{s^2}\,{x^3} + {s^3}\,{x^3}\nn
\eer
\[ J^{(4)}_{tZ} = 16\,{m_W^4}\,\left( 4\,{m_W^2} -
           2\,{m_Z^2} - s - 4\,{m_W^2}\,x + s\,x
            \right)
\]
\[
C^{(4)}_{t\gamma} = C^{(4)}_{tZ} \quad
J^{(4)}_{t\gamma} = J^{(4)}_{tZ}|_{m_Z=0}
\]
Higgs boson amplitude (\ref{m4h})
\[\M^{(4)}_H = g^2\,\frac{C^{(4)}_H}{J^{(4)}_H}\]
\ber
C^{(4)}_H &=& 32\,{m_H^2}\,{m_W^4} - 32\,{m_W^6} -
       24\,{m_H^2}\,{m_W^2}\,s + 24\,{m_W^4}\,s\nn\\ &+&
       5\,{m_H^2}\,{s^2} - 8\,{m_W^2}\,{s^2} + {s^3} +
       32\,{m_W^6}\,x + 8\,{m_H^2}\,{m_W^2}\,s\,x\nn\\ &-&
       40\,{m_W^4}\,s\,x - 2\,{m_H^2}\,{s^2}\,x +
       8\,{m_W^2}\,{s^2}\,x + {m_H^2}\,{s^2}\,{x^2} -
       {s^3}\,{x^2}\nn
\eer
\[J^{(4)}_H =
   8\,{m_W^2}\,\left( {m_H^2} - s \right) \,
     \left( 2\,{m_H^2} - 4\,{m_W^2} + s +
       4\,{m_W^2}\,x - s\,x \right)
\]
\end{appendix}

\end{document}